\documentclass[useAMS,usenatbib]{mnras}
\usepackage{times}
\usepackage{graphics,epsfig}
\usepackage{graphicx,amsmath,amsopn,multicol}
\usepackage{amssymb}
\usepackage{txfonts}
\usepackage{psfrag}
\usepackage{setspace}
\usepackage{amsfonts}                                                                              
\usepackage{multirow}
\expandafter\def\csname ver@subfig.sty\endcsname{}
\usepackage{placeins}
\usepackage{subfig}
\usepackage{mathrsfs}
\usepackage[utf8]{inputenc}
\usepackage{setspace}
\usepackage{mathtools}
\usepackage{grffile}
\usepackage{amsmath}
\usepackage{longtable}
\usepackage{graphics}
\usepackage{verbatim}
\usepackage[usenames]{color}
\usepackage{float}
\usepackage{bm}
\usepackage{epstopdf}

\newcommand{\U}{\mathbfit{U}}
\newcommand{\Ug}{\mathbfit{U}_{\rm g}}
\newcommand{\Ui}{\mathbfit{U}_{\rm i}}
\newcommand{\Un}{\mathbfit{U}^{'}}
\newcommand{\W}{\textit{W} (\bm{\theta})}
\newcommand{\A}{\textit{A} (\bm{\theta})}
\newcommand{\R}{\textit{R} (\bm{\theta})}

\newcommand{\tv}{\bm{\theta}}
\newcommand{\V}{\mathcal{V}}
\newcommand{\sig}{\mathcal{S}}
\newcommand{\N}{\mathcal{N}}
\newcommand{\T}{\delta \textit{T} (\bm{\theta})}
\newcommand{\Ts}{\delta \textit{T}_{s} (\bm{\theta})}

\title[Study of Kepler SNR using TGE]{A study of Kepler supernova remnant: angular power spectrum estimation from radio frequency data}
\author[Saha et al.]{Preetha Saha,$^{1}$\thanks{E-mail: preetha@phy.iitkgp.ernet.in} Somnath Bharadwaj,$^{1}$ Nirupam Roy,$^{2}$ Samir Choudhuri $^{3}$ 
\newauthor Debatri Chattopadhyay $^{4}$\\~\\ 
$^{1}$ Department of Physics and Centre for Theoretical Studies, Indian Institute of Technology, Kharagpur 721302, India\\
$^{2}$ Department of Physics, Indian Institute of Science, Bangalore 560012, India\\
$^{3}$ National Centre for Radio Astrophysics, Tata Institute of Fundamental Research, Pune University campus, Pune 411 007, India\\
$^{4}$ Centre for Astrophysics and Supercomputing, Swinburne Institute of Technology, John St, Hawthorn, Victoria 3122, Australia}
\begin{document}

\date{Accepted yyyy month dd. Received yyyy month dd; in original form yyyy month dd}

\pagerange{\pageref{firstpage}--\pageref{lastpage}} 
\pubyear{2018}

\maketitle

\label{firstpage}

\begin{abstract}
Supernova remnants (SNRs) have a variety of overall morphology as well as rich structures over a wide range of scales. Quantitative study of these structures can potentially reveal fluctuations of density and magnetic field originating from the interaction with ambient medium and turbulence in the expanding ejecta. We have used $1.5$GHz (L band) and $5$GHz (C band) VLA data to estimate the angular power spectrum $C_{\ell}$ of the synchrotron emission fluctuations of the Kepler SNR. This is done using the novel, visibility based, Tapered Gridded Estimator of $C_{\ell}$. We have found that, for $\ell = (1.9 - 6.9) \times 10^{4}$, the power spectrum is a broken power law with a break at $\ell = 3.3 \times 10^{4}$, and power law index of $-2.84\pm 0.07$ and $-4.39\pm 0.04$ before and after the break respectively. The slope $-2.84$ is consistent with 2D Kolmogorov turbulence and earlier measurements for the Tycho SNR. We interpret the break  to be related to the shell thickness of the SNR ($0.35 
$ pc) which approximately  matches 
$\ell = 3.3 \times 10^{4}$ (i.e., $0.48$ pc). However, for $\ell > 6.9 \times 10^{4}$, the estimated $C_{\ell}$ of L band is likely to have dominant contribution from the foregrounds while for C band the power law slope $-3.07\pm 0.02$  is roughly consistent with $3$D  Kolmogorov turbulence like that observed at large $\ell$ for Cas A and  Crab SNRs. 

\end{abstract} 


\section{Introduction} 
A supernova explosion marks the end stage of the life cycle of a massive star. The huge amount of kinetic energy released during the explosion, moves through the surrounding interstellar medium (ISM) in the form of a shock wave. As the shock travels ahead of the ejected stellar material, it collects materials from the ISM through heating and compression. When the mass swept up by the shock becomes almost comparable to that of the ejecta, the remnant enters into its adiabatic phase. This is generally characterized by a spherically symmetric model which assumes an isotropic explosion in a homogeneous ambient medium. This simplistic assumption leads to a self-similar solution which was given independently by  \cite{sedov1946} and \cite{Taylor1949}, for any strong point-like explosion. However, the one-dimensional self-similarity does not remain valid for other phases of evolution due to the interaction of the ejecta with the inhomogeneities in the ISM \citep{Chevalier1977}.

Besides the temporal evolution, at any epoch a supernova remnant (SNR) exhibits a variety of rich and complex structure across a wide range of length scale and frequency of observation. At radio wavelengths, the dominant contribution to the SNR emission comes from the non-thermal synchrotron radiation, emitted by relativistic electrons spiralling in the magnetic field.  A plausible hypothesis \citep{Gull1973a} is that the interaction between the ISM and the ejecta amounts to a convective instability, which makes sufficient turbulent energy available to account for the observed synchrotron radio emission. It is difficult to identify each underlying source of turbulence and precisely quantify its nature. However, the statistical quantification of the observed intensity fluctuations in the synchrotron radiation is expected to reveal interesting information about the fine spatial structures in the remnant.

\cite{roy09} have carried out a power spectrum analysis of the intensity fluctuations for two supernova remnants, Cas A and Crab. The power spectrum was estimated directly from the visibilities measured in the radio interferometric observations from Very Large Array (VLA) at frequencies $~1.5$ GHz (L band) and $~5$ GHz (C band). The estimator correlated pairs of visibilities (see Bare Estimator of \cite{Choudhuri2014}) and the resultant power spectrums of the SNRs were found to follow approximately a power law of index $-3.24\pm 0.03$. A break in power law from $-3.2$ to $-2.2$ was noted for Cas A owing to the shell-type structure of the remnant. Moreover, no such change in the power law of the power spectrum was observed for filled-type geometry of Crab. These power law power spectrums were shown to be roughly consistent with MHD turbulence in the synchrotron emitting plasma. For Tycho SNR, a Kolmogorov-like magnetic energy spectrum has been reported by analysing the spatial two-point correlation function of synchrotron intensities \citep{shimoda2018}. 

In this paper, we estimate the angular power spectrum of the Kepler SNR using two different VLA archival visibility data observed in the L  and C bands respectively. We use an improved estimator Tapered Gridded Estimator (TGE) \citep{Choudhuri2016b} which helps in computing the angular power spectrum directly from the visibilities by efficiently gridding the data, thereby reducing the computational time. This gridding property is not particularly important here since we handle a small amount of data. The estimator gives an added advantage of suppressing the residual point source contamination to a large extent by tapering the primary beam response through an appropriate convolution in the visibility domain. We also present the angular power spectrum estimates of Cas A and Crab as consistency checks for the estimator and to verify the results reported in \cite{roy09}.

Kepler's supernova (SN1604) was first recorded by Johannes Kepler. $5$GHz results at Cambridge and at Owens valley, revealed that Kepler SNR has a shell structure of $170\arcsec$ diameter, almost circular, but with irregular brightness distribution i.e brightest in the North, with a gap in the South and a peculiar dent in the East and the remnant is quite similar to Tycho's SNR  in term of its age, size, flux density and radio structure \citep{Gull1975}. Kepler SNR is estimated to be located at a distance of $(4.8-6.4)$ kpc \citep{ReyGoss99} using VLA H\,{\sc i} observations of the remnant. The remnant is believed to be a result of type Ia explosion. From deep Chandra observations, \cite{Reynolds2007} confirmed the high Fe/O ratio and found no evidence for a neutron star, further supporting the Type Ia origin. \cite{Patnaude} claimed that Kepler SNR was likely to be a $91$T-like event that produced nearly $1M_{\sun}$ $^{56}$Ni, by comparing the observed X-ray spectrum with simulated spectra based on their hydrodynamic modeling. By simply comparing the X-ray spectra, \cite{Katsuda2015} found that line intensity ratios of iron-group elements (IGE) to intermediate-mass elements (IME) for Kepler SNR and SNR $0509-67.5$ are much higher than those for Tycho SNR and therefore argued that Kepler SNR is the product of an overlumious Type Ia SNe.

We present the estimates of the angular power spectrum of the observed intensity fluctuations from the VLA archival data of the Kepler SNR using TGE. The details of the archival data used here and its analysis technique are briefly outlined in Section \ref{sec:data}. The methodology of the power spectrum estimation and its error are described in Section \ref{sec:psest}.  We discuss the results of Cas A, Crab SNR in Section \ref{sec:CasA&Crab}. We present the results of Kepler SNR in Section \ref{sec:results}, followed by its discussion and conclusion in Section \ref{sec:conclusion}.

\section{Data}
\label{sec:data}
\begin{figure*}
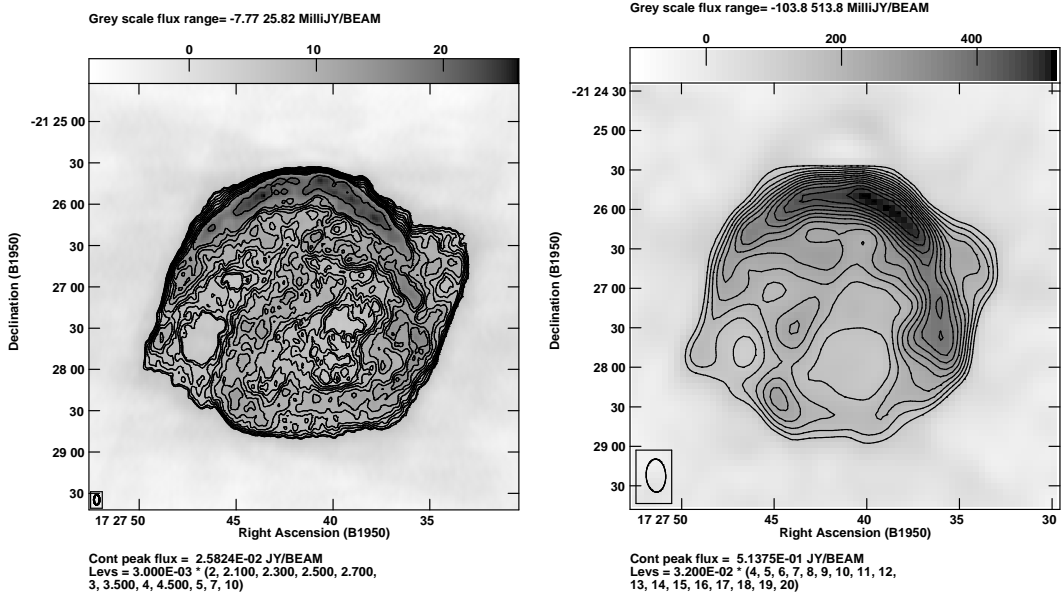

 \begin{center}
 \includegraphics[scale=0.37]{kepCC11_3_1.eps}	 \includegraphics[scale=0.37]{kepLC11_2.eps}
 \caption{Total intensity contour plots of the Kepler SNR obtained using AIPS from the calibrated visibility data set of C (left)   and L (right)  bands respectively. Only the VLA C configuration data were used for both the plots.}
 \label{fig:AIPSkeplerCC}
 \end{center}
\end{figure*}

We use multi-configuration archival data (details given in Table \ref{tab:Keplerdata}) of VLA observations of Kepler SNR in the L  and C  frequency bands. Here we reduce the single frequency and line-free channel data of Kepler SNR in each configuration using the standard tasks (\cite{tora.book}, \cite{isra.book}) in classic AIPS\footnote{NRAO Astrophysical Image Processing System, a commonly used software for radio data processing}. The bad visibility points are flagged following which the flux density scale and instrumental phase are calibrated. The calibrated visibility data of the target source is subsequently separated from the multi-source $\textit{uv}$ data of a given configuration using the task SPLIT. We convert the calibrated visibilities of the SNR into a CLEANed radio image (shown in Figure \ref{fig:AIPSkeplerCC}) to verify the flagging and calibration.  The angular diameter of the remnant is inferred from the image along the north-south and east-west directions by identifying the boundary of the remnant. The approximate mean diameter evaluated for Kepler SNR are $3.70\arcmin$ and $3.30\arcmin$ in L and C bands respectively.  

\begin{table}
		\centering
		\begin{tabular}{p{0.5cm}p{0.5cm}p{2cm}p{1cm}p{0.7cm}p{1.5cm}}
			\hline
			Band & Array & Observation date & Bandwidth (MHz) & ToS (mins) & Central frequency (GHz)\\ \hline
			C & A & 2004 December 17 & 25 & 318 & IF1 : 4.7149, IF2: 4.9851\\ \hline
			C & B & 2005 March 8 & 50 & 245 &  IF1 : 4.7149, IF2: 4.9851\\ \hline
			C & C & 2004 May 6 & 50 & 165 & IF1 : 4.7149, IF2: 4.9851\\ \hline
			L & A & 2004 December 19 & 25 & 319 & IF1 : 1.4649, IF2: 1.3649\\ \hline
			L & B & 2005 April 27 & 50 & 216 & IF1 : 1.4649, IF2: 1.3649\\ \hline
			L & C & 2004 May 6 & 50 & 165 & IF1 : 1.2851, IF2: 1.4649\\ \hline
		\end{tabular}
		\caption{Specifications of the VLA archival data for Kepler  SNR bearing project code AD$498$. The fifth column ToS stands for the total on-source time.}
		\label{tab:Keplerdata}
	\end{table}

\section{Methodology}
\label{sec:psest}
\subsection{Power spectrum estimation using TGE} 

We apply the visibility based TGE (details in \cite{Choudhuri2014,Choudhuri2016b}) to the calibrated visibility data of the SNR to estimate the angular power spectrum $C_{\ell}$ of the measured brightness temperature fluctuations. Each measured visibility  $\V_{i}$ is assumed to be the sum of two contributions namely the sky signal 
 $\sig(\Ui)$  and a system noise contribution $\N_{i}$
 \begin{equation}
 \V_{i}=\sig(\Ui)+\N_{i} \,.
 \label{eq:v1}
\end{equation} 

The signal component $\sig(\Ui)$ is the Fourier transform of the product of the telescope's primary beam pattern $\A$ and the  brightness temperature fluctuations $\T$ on the sky (eg.  eq. ($10.2.10$) of  \citealt{gmrt.book})
\begin{equation}
    \label{s1}
     \sig(\Ui)=Q_{\nu} \int d^2 \tv \, e^{2\pi i \Ui \cdot\tv} \, \A \, \T 
\end{equation}
where $\tv$  is a 2D vector in the plane of the sky with $\theta=\mid \tv \mid$. 
Here  $Q_{\nu}=2 k_B/\lambda^2$ is the conversion factor from brightness temperature  to specific intensity in the Rayleigh Jeans limit and $k_B$ is the Boltzmann constant. Eq.~(\ref{s1}) can also be written in terms of a convolution as
\begin{equation}
    \label{s2}
    \sig(\Ui)=Q_{\nu} \int d^2 {\U} \, \tilde{a}(\Ui-\U) \, \Delta \tilde{T} (\U)
\end{equation}
where $\tilde{a}(\U)$ and $\Delta \tilde{T}(\U)$ are  the  Fourier transforms of $\A$ and $\T$ respectively.
Here we assume that the observed sky signal $\T$ is a particular realization of a statistically homogeneous and isotropic Gaussian random field whose statistical properties are completely quantified by  the  angular power spectrum  $C_{\ell}$   defined using 
\begin{equation}
    \label {C_l}
    \langle  \Delta \tilde{T}(\U) \, \Delta \tilde{T}^{*}(\Un) \rangle  = \delta^{2}_{D}(\U-\Un)\,C_{2\pi U} 
\end{equation}
where the angular brackets denotes an ensemble average over different  realizations of the random field $\T$. The estimator here aims to determine  $C_{\ell}$ from $\V_i$ the measured visibilities.

The primary beam pattern $\A$ quantifies how the individual antenna responds to signals from different directions $\tv$ in the sky. The primary beam pattern $\A$ typically has a full width at half maxima $(\theta_{FWHM})$  of $\sim \lambda/D$ where $D$ is the antenna diameter, and  for example  $\theta_{FWHM}\sim30^{\arcmin}$ for VLA in the L band. The target source one wishes to observe is usually smaller than the angular extent of $\A$, and for example it is $\sim 3^{\arcmin}$ for the Kepler SNR. In addition to the sky signal from the target source, the measured visibilities $ \V_{i}$ may have significant contributions from other sources which lie within the angular extent of $\A$.  It is therefore desirable to restrict the sky response 
to a small region around the target source and thereby avoid contributions from extraneous sources.
The TGE achieves this by tapering the sky response with a suitably chosen window function $\W$. 
Here we have used $\W =e^{-\theta^{2}/\theta_{w}^{2}}$ with   $\theta_{w}=0.6 \, f \, \theta_{FWHM}$
where the parameter $0<f\le 1$ controls the amount of tapering. The sky response gets narrower as the value of $f$ is reduced. This tapering  is introduced by convolving the measured visibilities with $\tilde{w}(\U)$ which is the Fourier transfrom of $\W$.
This  is implemented by dividing the \textit{uv} plane into a rectangular grid and calculating the convolved visibility (eq. \ref{V_con&grid}) at each grid point $g$ \begin{equation}
 \label{V_con&grid}
\V_{cg}=\sum_{i} \tilde{w}(\Ug-\Ui)\V_{i}
\end{equation} 
where $\Ug$ refers to the baseline  corresponding to the grid point $g$. The size of the rectangular $\textit{uv}$ grid is set by the choice of $U_{max}$ and $U_{min}$ such that each grid can incorporate the visibility data of both the largest configuration and the shortest configuration of VLA. We have chosen a grid spacing within which the convolution in eq. \eqref{V_con&grid} is well represented. The uniform grid size and spacing allows us to collapse the visibility data of the different VLA configurations into a single grid. Note that  the two observing bands L and C of VLA were analyzed separately. 
The gridded and convolved visibility $\V_{cg}$ in eq. (\ref{V_con&grid}) can be expressed  in terms of the Fourier components $\Delta \tilde{T}(\U)$ as 
\begin{equation}
\label{Vcg}
    \V_{cg}=Q_{\nu} \int d^2 \U \, \tilde{K}(\Ug-\U) \, \Delta \tilde{T}(\U) + \sum_{i} \tilde{w}(\Ug-\Ui)\N_{i}
\end{equation}
where 
\begin{equation}
\label{kernel}
    \tilde{K}(\Ug-\U) = \int d^{2} \Un \, \tilde{w}(\Ug-\Un)\, B(\Un) \, \tilde{a}(\Un-\U)
\end{equation}
 is the effective gridding kernel and 
 \begin{equation}
     B(\U)=\sum_{i} \delta^{2}_{D} (\U-\Ui)
 \end{equation}
 is the baseline sampling in the \textit{uv} plane. The self correlation of the gridded and convolved visibilities  is given by
  \begin{equation}
  \label{Vcg_selfco}
  |\V_{cg}|^{2}= Q_{\nu}^{2}\int d^{2}{\U}\,|\tilde{K}(\Ug-\U)|^{2} \, C_{2\pi U}+ \sum_{i} |\tilde{w}(\Ug-\Ui)|^{2}|\N_{i}|^{2} \,.
  \end{equation}
We expect the function $\mid \tilde{K}(\U)\mid^{2}$ to be peaked around $\mid \U \mid=0$ and have a narrow width in comparison to $C_{2\pi U}$. At large baselines $\mid \U \mid$, or equivalently at large angular multipoles $\ell$, 
the convolution in eq. (\ref{Vcg_selfco}) can be  approximated as
  \begin{equation}
    \label{Vcg_selfco_approx}
   |\V_{cg}|^{2}=\bigg [Q_{\nu}^{2}\int d^{2}{\U}\,|\tilde{K}(\Ug-\U)|^{2}\bigg]\,C_{2\pi U}+ \sum_{i} |\tilde{w}(\Ug-\Ui)|^{2}|\N_{i}|^{2}
    \end{equation}
under the assumption that $C_{2\pi U}$ is nearly constant across the width of $\tilde{K}(\Ug-\U)$.   The estimator is then defined as
   \begin{equation}
   \label{eqn:TGE_V_m}
   \hat{E}_{g}=(M_{g})^{-1}\times\bigg(|\V_{cg}|^{2}-\sum_{i} |\tilde{w}(\Ug-\Ui)|^{2}|\V_{i}|^{2}\bigg)
   \end{equation}
   where $M_{g}$ is a normalization constant and $ \langle\hat{E}_{g}\rangle$ gives an unbiased estimate of the angular power spectrum $C_{\ell}$ at the angular multipole $\ell_{g}=2 \pi \mid \Ug \mid$.  The values estimated at the individual grid points were averaged in circular bins of equal logarithmic interval in  $\ell$, and we present the bin averaged  values of the angular power spectrum  $C_{\ell}$. We refer to this estimated angular power spectrum as $C^E_{\ell}$. 

We have used simulations of the observed visibilities to determine the normalization constant $M_g$. Similar simulations were also used for  the error estimates  presented in Section \ref{sec:error} of this paper. 
We next discuss how we have simulated the visibilites corresponding to an input model angular power spectrum  $C^M_{\ell}$ for the sky signal.   For simulating the visibilities we assume that the brightness temperature fluctuations $\T$ on the sky are realizations of a statistically homogeneous and isotropic Gaussian random field  whose statistical properties are completely determined by the angular power spectrum. We generate the Fourier components of the brightness temperature fluctuations corresponding to the input model angular power spectrum $C_{\ell}^{M}$ using 
     \begin{equation}
     \label{I_fourier_fluc}
     \Delta \tilde{T}(\U)=\sqrt{ \frac{\Omega C_{\ell}^{M}}{2}}[x(\U)+iy(\U)]
     \end{equation}
where $\Omega$ is the total solid angle of the simulation $\ell=2 \pi |\U| $, and $x(\U)$ ,$y(\U)$ are independent Gaussian random variables with zero mean and unit variance. The simulations were carried out  on  synthesized grids of size $4096\times 4096$ and $2048\times 2048$ with  resolution $0.005\arcmin$ and $0.02\arcmin$  for the C and L bands of VLA respectively. For each sampled baseline $\U_i$,
the sky signal contribution $\sig(\U_i)$ to the simulated visibility $\V_i$ is computed from the Fourier transform
of the product (eq.~\ref{s1}) of the quantity $Q_{\nu}\,\T$ with the telescope's primary beam $\A$. In addition to the sky signal, the visibilities also have a system noise contribution $\N_i$ \, (eq. \ref{eq:v1}). For a single polarization, we have modelled both the real and imaginary components of $\N_i$ as Gaussian random variables with zero mean and variance $\sigma_N^2$ whose value we have specified later.  The simulated visibilities have  the same baseline distribution as the actual measured data. 
The simulations closely follow   the methodology 
presented in \citet{2017NewA...57...94C} to which the reader is referred for further details. 

The value of the normalization constant $M_g$ is calculated using
   \begin{equation}
     M_{g}=\bigg \langle \, \big (|\V_{cg}|^{2}-\sum_{i} |\tilde{w}(\Ug-\Ui)|^{2}|\V_{i}|^{2} \big )\, \bigg \rangle UAPS
   \end{equation}
   where angular brackets denote an ensemble average over different realizations of the simulated visibilities corresponding to an unit angular power spectrum (UAPS) for which $C^M_{\ell}=C^{UAPS}_{\ell}=1$. Note that the UAPS simulations only have the sky signal contribution, and there is no system noise.
Here we have averaged over $200$ independent realizations of the UAPS simulation to reduce the statistical uncertainty in the estimated $M_{g}$. 

Various approximations, including that for the convolution  in eq.~(\ref{Vcg_selfco}), 
have been extensively tested in \citet{Choudhuri2014}. The fact that the TGE is able to faithfully recover $C_{\ell}$ from the visibility data even in the presence of sparse sampling of the baselines  has been validated using simulations  in \citet{Choudhuri2016b}. A detailed study \citep{Choudhuri2016a} has demonstrated, using simulations, that the TGE is able to suppress the contribution from extraneous sources (foregrounds) in the outer regions of the telescope's field of view (FoV) while
faithfully estimating $C_{\ell}$ from the visibility data. The same has been recently demonstrated using $325 \, {\rm MHz}$ uGMRT 
 observations \citep{Chakraborty2019} of the diffuse Galactic synchrotron emission (DGSE).

\subsection{Interpreting $C^E_{\ell}$}
In the simplest picture, we may think of a SNR as an expanding spherically symmetric shock propagating in an uniform ISM, as was originally considered by \citet{sedov1946} and \citet{Taylor1949}. The resulting brightness temperature fluctuations on the sky can be represented as 
\begin{equation}
\label{eq:snr.a}
    \T=\R\,\bar{T}_{s}
\end{equation}
where $\R$ is a dimensionless profile function with $R(0)=1$ at the center of the SNR and $\R=0$ for $\theta > \theta_r$, i.e. it is zero beyond the angular extent $\theta_r$ of the SNR and $\bar{T}_{s}$ is the brightness temperature at the center of the SNR. The profile function $\R$ captures the angular profile of the remnant. We however see (Figure \ref{fig:AIPSkeplerCC}, also in \citet{roy09}) that the SNR exhibits structures spanning nearly the entire range of accessible angular scales i.e from $\theta_r$ to the angular resolution of the observations. We interpret  these fluctuations as arising from MHD turbulence in the SNR.

A statistical interpretation of turbulence is well accepted in the literature (\citet{taylor1935statistical}, \citet{monin1971statistical}).
 Considering the turbulent ISM of our Galaxy,  \citet{CM_V1952} have modelled this in terms of a continuous matter distribution  with a mean and a fluctuating component. The fluctuating component was assumed to be a statistically homogeneous and isotropic random field whose statistical properties were described by the two-point correlation function. In line with this work, we model the brightness temperature fluctuation $\T$ of the radiation received from the SNR through
\begin{equation}
\T=\R [\bar{T}_{s} + \Ts]
\label{eq:snr.b}
\end{equation}
where  we have assumed the fluctuating component  $\Ts$ to be the outcome of a statistically homogeneous and isotropic Gaussian random process (presumably turbulence) whose statistical properties are completely specified by the angular power spectrum $C_{\ell}$. Note that $C_{\ell}$ is the Fourier transform of the two point correlation function. Here
both $\bar{T}_{s}$ and $\Ts$ are modulated by a dimensionless profile function $\R$  which incorporates the radial profile of the SNR and cuts off the emission beyond the finite angular extent of the SNR.  Similar models have also been used in earlier  works \citep{begum2006,Dutta2008,dutta2009,Dutta2013} which have estimated and interpreted the angular power spectrum of the HI 21-cm emission from several external galaxies.

The angular power spectrum $C^E_{\ell}$ estimated from $\T$ is related to $C_{\ell}$ 
(which corresponds to $\Ts$  )  through a convolution 
\begin{equation}
C^E_{2 \pi \mid \U \mid} =\int d^2 \U^{'} \, \mid \tilde{r}(\U-\U^{'}) \mid^2  \, C_{2 \pi \mid \U^{'} \mid} 
\label{eq:snr.b}
\end{equation}
where $\tilde{r}(\U)$ is the Fourier transform of $\R$.  Considering a power law of the form $C_{\ell} \propto A \, \ell^{\beta}$ with a negative power law index $(\beta <0)$, at large $\ell$ $(\gg \ell_m)$ the convolution is well approximated by 
\begin{equation}
C^E_{2 \pi \mid \U \mid} =\left[ \int d^2 \U^{'} \, \mid \tilde{r}(\U^{'}) \mid^2 \right]
\, C_{2 \pi \mid \U \mid} 
\label{eq:snr.c}
\end{equation}
 where the estimated $C^E_{\ell}$ has the same slope as $C_{\ell}$, and the two differ only differ by a proportionality constant $\left[ \int d^2 \tv \, \mid \R \mid^2 \right]=\left[ \int d^2 \U^{'} \, \mid \tilde{r}(\U^{'}) \mid^2 \right]$ \citep{dutta2009}. However, at small $\ell$ $(\le \ell_m)$ the convolution 
(eq. \ref{eq:snr.b}) introduces a break at $\ell_m$. The shape of $C^E_{\ell}$ differs significantly from that of $C_{\ell}$ at $\ell \le \ell_m$, and $C^E_{\ell}$ flattens out in this range.  The value of $\ell_m$ is inversely proportional to the angular extent of the SNR, the exact value of $\ell_m$ however depends on the slope $\beta$ and the shape of the profile function $\R$. In the $\ell$ range  $\ell \gg \ell_m$ the estimated $C^E_{\ell}$ is proportional to $C_{\ell}$, and the estimated $C^E_{\ell}$ values  can be used to determine the slope of $C_{\ell}$. This $\ell$ range can also be used to determine the amplitude of $C_{\ell}$ provided we have  a precise model for $\R$,  however we have not attempted this here. 

In the present work, we have modelled the angular profile of the SNR   as a  Gaussian of the form  $\R =e^{-{\theta^{2}/\theta_{r}^{2}}}$. For each data the value of $\theta_{r}$ was chosen so as to correctly reproduce the break $\ell_m$ in $C^E_{\ell}$ (as discussed later). For each data, the chosen value of $\theta_{r}$ was also found to be roughly consistent with the angular extent of the SNR estimated from the  image made with the corresponding data. We have used the values   $\theta_{r}= 0.8\arcmin, 1.2\arcmin,1.3\arcmin, 0.7\arcmin$ for Kepler SNR in C and L bands, Cas A and Crab SNR respectively.

\subsection{Error estimation}
\label{sec:error}
The statistical fluctuations inherent to the sky signal $\sig(\U_i)$ as well as the system noise contribution $\N_i$ both contribute to statistical errors in the estimated angular power spectrum $C^E_{\ell}$. We have used simulations to estimate these errors.       
We have used simulations to generate several  statistically independent realizations of the visibilities $\V_i$  for which the average   $C_{\ell}$ matches $C^E_{\ell}$, the variance determined from the multiple realizations of the  simulation is used to 
estimate the statistical errors $\delta C_{\ell}$ in the estimated $C^E_{\ell}$. The simulations were carried out using the methodology outlined earlier  and for which the details are presented in 
\citet{2017NewA...57...94C}.

As mentioned earlier, the slope of $C^E_{\ell}$ matches that of the sky signal $C_{\ell}$ at $\ell \gg \ell_m$
the amplitude however differs. The convolution (eq. \ref{eq:snr.b}) causes $C^E_{\ell}$ to differ from 
 $C_{\ell}$ at $\ell \le \ell_m$. To  simulate the sky signal we have assumed that 
 $C^M_{\ell}=B^{-1} \, C^E_{\ell}$ for $\ell \ge \ell_m$ and we have applied a spline interpolation to
 the estimated  $C^E_{\ell}$ to obtain a continuous function of $\ell$. We have fitted a power law to 
 $C^E_{\ell}$ at the vicinity of $\ell \approx \ell_m$ (with $\ell \ge \ell_m$) and we have extrapolated this   for $C^M_{\ell}$  at $\ell < \ell_m$.  The amplitude of $C^M_{\ell}$ is  $B^{-1}= \left[ \int d^2 \tv \, \mid \R \mid^2 \right]$, however the actual form of $\R$ is not precisely known and we have set the value of $B^{-1}$ so that $C_{\ell}$ estimated from the simulated visibilities matches $C^E_{\ell}$ estimated from the actual data. We find that the values of $B^{-1}$ used here satisfy  $B^{-1} \approx \left[ \int d^2 \tv \, \mid \R \mid^2 \right]$ which is expected from our analysis. For the system noise contribution we have used 
 
\begin{equation}
\label{sys_noise_formula}
\sigma_N=\frac{SEFD}{\eta_{c}\sqrt{2t_{int}\Delta \nu}}
\end{equation}
where SEFD, $\eta_{c}$ \footnote{https://science.nrao.edu/facilities/vla/docs/manuals/oss/performance/sensitivity} 
are the system equivalent flux density (Jy) and the correlator efficiency respectively, $t_{int}$ is the integration time per visibility in seconds and  $\Delta \nu$ is the channel width in Hz. The values of 
$t_{int}$, $\Delta \nu$ and  $\sigma_N$ which we have used here have been tabulated in Table \ref{tab:Keplernoise} (for Kepler SNR) and  
Table \ref{tab:CasCrabnoise} (for Cas A and Crab SNR). Note that these values are different for the different sources, frequency bands and VLA configurations. 

In this work we have generated $10$ statistically independent realization of the simulated visibilities. The resulting estimates of $C_{\ell}$ were used to determine  the error $\delta C_{\ell}$ for the measured $C^E_{\ell}$.

\begin{table}
	\centering
	\begin{tabular}{|c|c|c|c|c|}
		\hline
		Band & Configuration & Channel width $\Delta \nu$ MHz & $t_{int}$ secs & $\sigma_N$ (mJy)\\ \hline
		L     &        A      &    15.625     &   3.3    &       65 \\ \hline
		L     &        B      &       37.5    &     5     &         34 \\ \hline
		L     &        C      &       50    &     10    &        21 \\ \hline
		C    &       A     &     15.625    &   3.3       &    53 \\ \hline
		C     &       B    &         37.5   &      5    &         28 \\ \hline
		C    &        C    &           50  &      10    &       17 \\ \hline
	\end{tabular}
	\caption{Channel width $\Delta \nu$, integration time $t_{int}$ and r.m.s. system noise $\sigma_N$ eq. (\ref{sys_noise_formula}) for Kepler SNR in A, B, C configurations of VLA L and C bands.}
	\label{tab:Keplernoise}
	\end{table}	
	\begin{table}
	\centering
	\begin{tabular}{|c|c|c|c|c|}
		\hline
		Source & Configuration & Channel width $\Delta \nu$ MHz & $t_{int}$ secs & $\sigma_N$ (mJy)\\ \hline
		Cas A     &        A      &    6.25     &   10   &       119 \\ \hline
		Cas A     &        B      &       12.5    &     10     &         85 \\ \hline
		Cas A     &        C      &       25    &     10    &        60 \\ \hline
		Cas A     &        D      &       25    &     10    &        60 \\ \hline
		Crab    &       A     &     50    &   10       &    37 \\ \hline
		Crab     &       B    &       25   &     10    &         53 \\ \hline
		Crab    &        C    &           25  &      10    &       53\\ \hline
		Crab    &        D    &           50  &      10    &       37\\ \hline
	\end{tabular}
	\caption{Channel width $\Delta \nu$, integration time $t_{int}$ and r.m.s. system noise $\sigma_N$ eq. (\ref{sys_noise_formula}) for Cas A and Crab SNR in A, B, C, D configurations of VLA C band}
	\label{tab:CasCrabnoise}
	\end{table}

\section{Cas A and Crab revisited}
\label{sec:CasA&Crab}
An earlier work  \cite{roy09} has analyzed the angular power spectrum of the Cas A and Crab SNRs. Their analysis was  however based on a different estimator (the ``Bare Estimator" of \cite{Choudhuri2014}) which uses the pairwise correlation of the individual visibilities to estimate $C_{\ell}$.  The TGE used here is computationally more efficient in that it deals with the gridded visibilities, it also allows the sky response to be tapered  to eliminate the contribution from extraneous 
sources located far away from the target source. In order to test if the results from the two different methodologies are consistent, and also to have $C_{\ell}$ measurements for 
Cas A, Crab and Kepler SNR using the same estimator, we have first revisited Cas A and Crab. 
 Here we re-analyze the calibrated VLA C band visibility data of Cas A and Crab SNR (details of the data in Table $1$ of \cite{roy09}) following the methodology described in Section \ref{sec:psest}. 

\begin{figure*}
 \begin{center}
  \psfrag{C}[top]{$\ell(\ell+1)\,C^E_\ell/2\pi$ \quad ${\rm K}^2$} \psfrag{ l}{} \psfrag{ l}[bottom]{$\ell$}  \psfrag{ K}{}
  \psfrag{10}{} 
 \psfrag{2}[bottom]{$10^2$} \psfrag{3}[bottom]{$10^3$} \psfrag{4}[bottom]{$10^4$} \psfrag{5}[bottom]{$10^5$} \psfrag{6}[bottom]{$10^6$} \psfrag{7}[bottom]{$10^7$}
 \psfrag{4}[right]{$10^4$} \psfrag{3}[right]{$10^3$} \psfrag{2}[right]{$10^2$} \psfrag{1}[right]{$10^1$} \psfrag{0}[right]{$10^0$} \psfrag{-1}[right]{$10^{-1}$} \psfrag{-2}[right]{$10^{-2}$} \psfrag{-3}{} \psfrag{-4}[right]{$10^{-4}$} \psfrag{-5}{} \psfrag{-6}[right]{$10^{-6}$} \psfrag{-7}{} \psfrag{-8}[right]{$10^{-8}$} \psfrag{-9}{} \psfrag{-10}[right]{$10^{-10}$} \psfrag{-11}{} \psfrag{-12}[right]{$10^{-12}$}
 \includegraphics[scale=0.30,angle=-90]{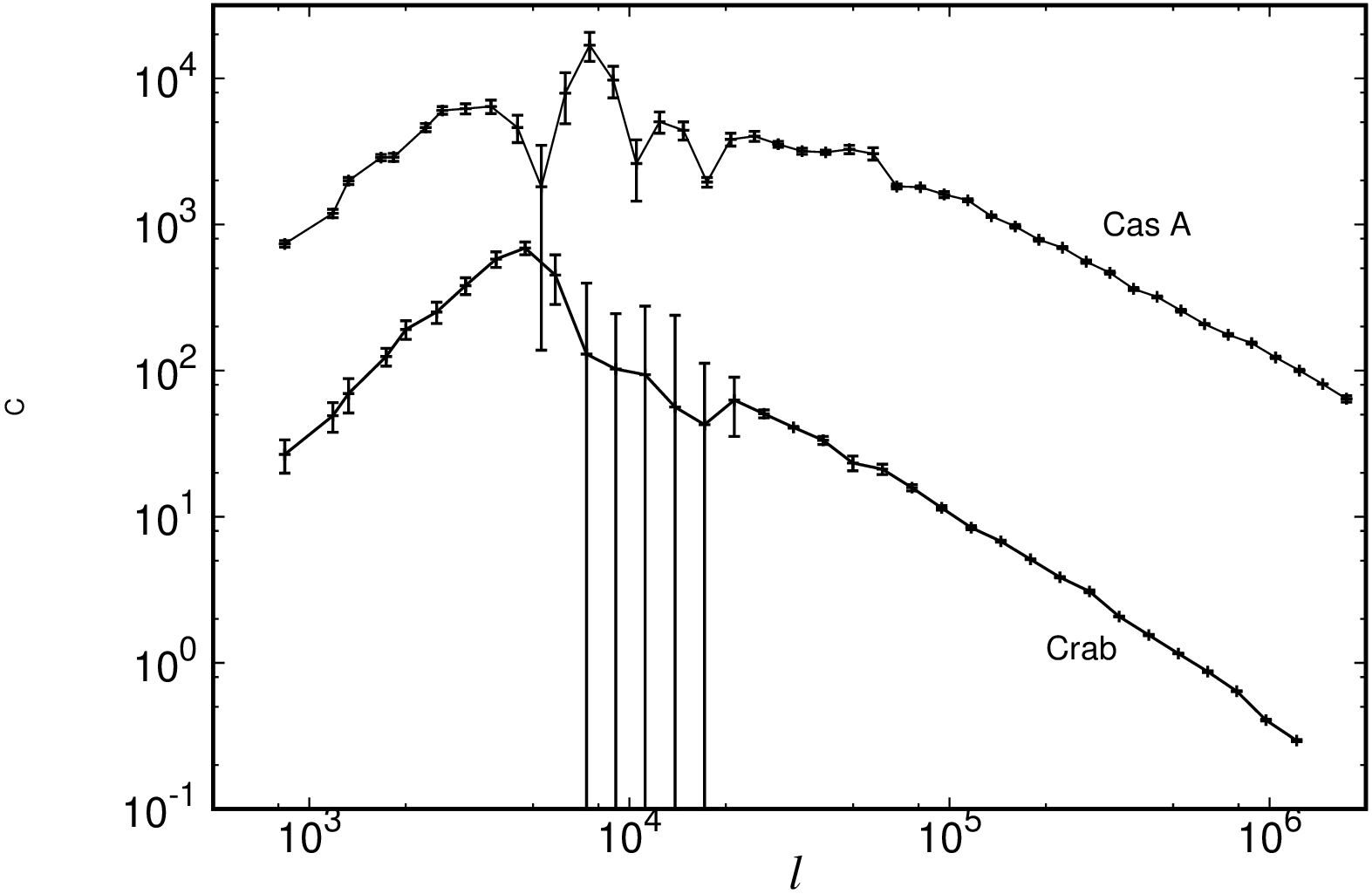}
 \hspace{0.5cm}
   \psfrag{l}[t]{$\ell$} \psfrag{  l}[b]{$\ell$}
   \psfrag{10}{} 
  \psfrag{2}[bottom]{$10^2$} \psfrag{3}[bottom]{$10^3$} \psfrag{4}[bottom]{$10^4$} \psfrag{5}[bottom]{$10^5$} \psfrag{6}[bottom]{$10^6$} \psfrag{7}[bottom]{$10^7$}
  \psfrag{4}[right]{$10^4$} \psfrag{3}[right]{$10^3$} \psfrag{2}[right]{$10^2$} \psfrag{1}[right]{$10^1$} \psfrag{0}[right]{$10^0$} \psfrag{-1}[right]{$10^{-1}$} \psfrag{-2}[right]{$10^{-2}$} \psfrag{-3}{} \psfrag{-4}[right]{$10^{-4}$} \psfrag{-5}{} \psfrag{-6}[right]{$10^{-6}$} \psfrag{-7}{} \psfrag{-8}[right]{$10^{-8}$} \psfrag{-9}{} \psfrag{-10}[right]{$10^{-10}$} \psfrag{-11}{} \psfrag{-12}[right]{$10^{-12}$}
 \includegraphics[scale=0.30,angle=-90]{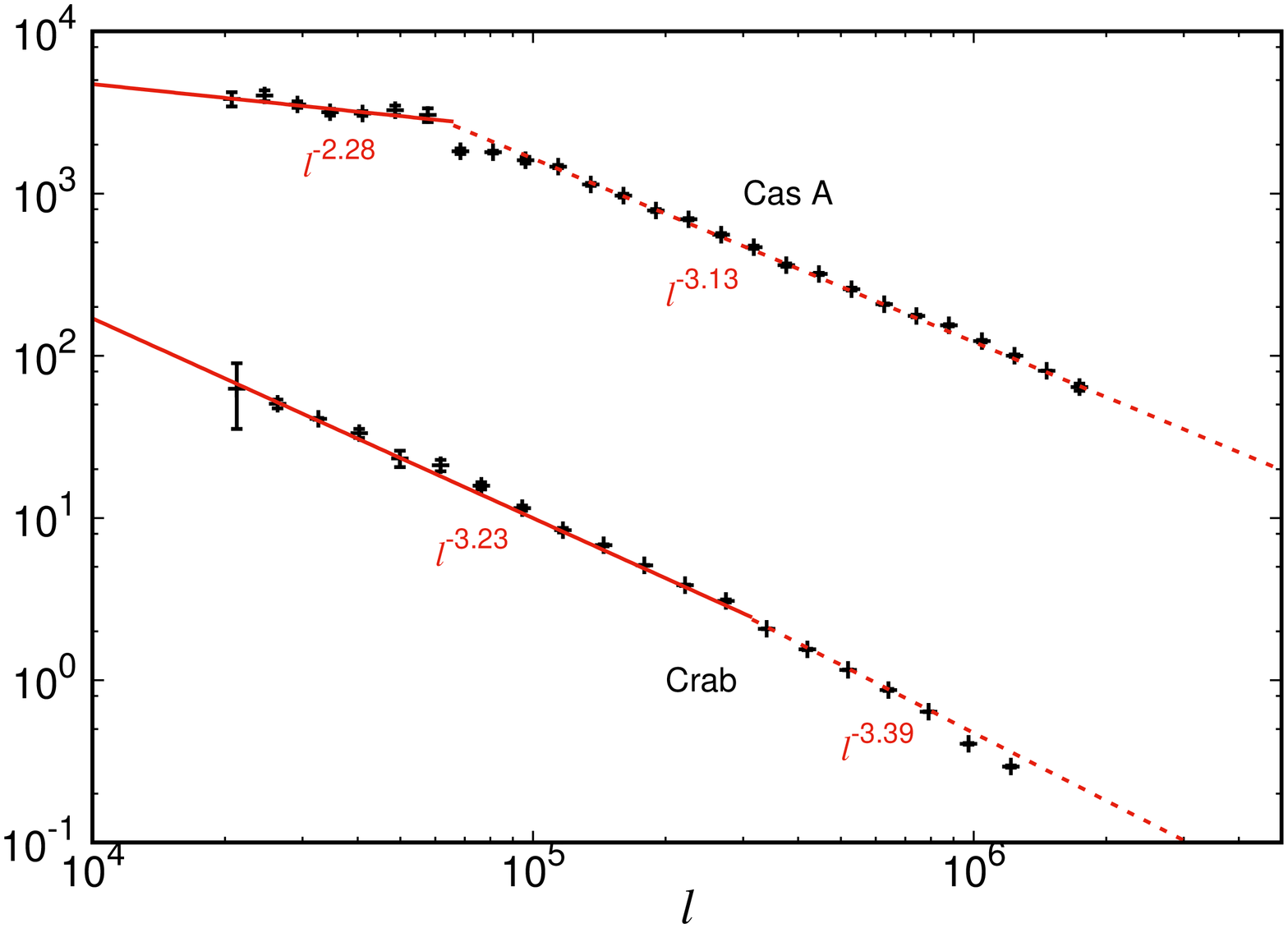}
 \caption{Estimated scaled  angular power spectrum    $\ell (\ell+1)\, C^E_{\ell}/(2 \pi) $ 
of  Cas A and Crab SNRs as a function of angular multipole $\ell$. The $\pm 1\sigma$ error bars are derived by simulating the observations of A,B,C,D configurations in VLA C ($5$GHz) band. The convolution dominated region for $\ell<10^{4}$ has not been shown in the right panel. The best fit lines $C^E_{\ell}\propto \ell^{\,\beta}$ are plotted with red solid and dotted lines in the right panel. The corresponding fit values of the  power-law index $\beta$ for the two $\ell$ ranges are shown in Table \ref{tab:CasAfit} \& Table \ref{tab:Crabfit} for Cas A and Crab SNR respectively.}
 \label{fig:CasACrab}
 \end{center}
 \end{figure*}

 The left panel of Figure \ref{fig:CasACrab} shows $C^E_{\ell}$ estimated using TGE for both the Cas A and Crab SNR. It may be noted that instead of $C^E_{\ell}$, throughout  the paper
 we have shown  the scaled angular power spectrum $\ell(\ell+1)\,C^E_{\ell}/2\pi$ which can be interpreted as the variance of the brightness temperature fluctuations at the  angular scale corresponding to $\ell$.
 The $\ell$ range shown here corresponds to the entire  baseline range of the observations. For both the SNRs  $C^E_{\ell}$ appears to be a power law with a negative power-law index at large $\ell$,.  The error bars of $C^E_{\ell}$ are not noticeable in this $\ell$ range. We see that  $C^E_{\ell}$ flattens out  at $\ell \leq \ell_m(= 10^4)$. This can  be attributed to the convolution (eq. \ref{eq:snr.b}) due to the finite angular size of the SNR. The error bars at small   $\ell$  are quite large  due to the sample variance.  
 
Considering $\ell > 10^4$, for both the SNRs we find that a single power law of the form $C=A\ell^{\beta}$ does not provide a good fit to this $\ell$ range. However, a better fit is obtained with broken power law where the slope has two different values  for  $\ell > \ell_b$ and $\ell < \ell_b$ respectively.  
 For each SNR  we have visually identified  an $\ell$  range 
within which $C^E_{\ell}$ appears to be single power law in $\ell$, and we have used  
$\chi^{2}$ minimization to determine the best fit values of the amplitude $A$ and power law index (slope) $\beta$.  The right panel of Figure \ref{fig:CasACrab} shows the best fit power laws obtained for the Cas A and Crab SNRs.  The $\ell$ range and the best fit power law indices are tabulated in Table~\ref{tab:CasAfit} and Table~\ref{tab:Crabfit} for Cas A and Crab SNRs respectively. 
 
We find that  for Cas A  $C^E_{\ell}$ shows a break at  $\ell_b =6.60 \times 10^{4}$ with $\beta=-2.28\pm 0.08$  and $\beta=-3.13\pm 0.01$ for $\ell < \ell_b$ and $\ell > \ell_b$ respectively. These findings are consistent
with the results of \cite{roy09} within the measurement errors bars.
The steepening of the power law at $\ell > \ell_b$ has been attributed to a transition from 2D to 3D MHD turbulence at scales smaller than the shell thickness of the Cas A SNR.  The large  value of the reduced $\chi^{2}$ 
for the fit at $\ell > \ell_b$ suggests that the errors in $C^E_{\ell}$ have possibly been  underestimated in this $\ell$ range.
 
 Considering Crab, we find a break at $\ell_b=3.14 \times 10^{5}$ 
 with $\beta=-3.23\pm 0.01$ and $\beta=-3.39\pm 0.01$ for $\ell < \ell_b$ and $\ell > \ell_b$ respectively. In contrast, \cite{roy09} found that   a single power law with $\beta=-3.24\pm0.03$ provides a good fit consistent with  MHD turbulence in the  filled-centre supernova remnant of Crab. The slope obtained at $\ell <  \ell_b$ in the present paper is consistent with the results of \cite{roy09}. However, it is  difficult to associate a morphological feature with the change of $0.16$ in the value of $\beta$ found at $\ell > \ell_b$ in the present paper. It may be possible that there is a gradual steepening of the power spectrum at small scales (large $\ell$), however it is not possible to say anything conclusive regarding this with the present data.   
 
 In conclusion of this section we note that the angular power spectrum obtained  using TGE are broadly consistent with the earlier results of  \cite{roy09}. 

 \begin{table}
 \centering
 \begin{tabular}{c|c|c|c|}
          		\hline
          		$\ell$ range & power index $\beta$ & No. of points $N$ & $\frac{\chi^{2}}{(N-2)}$\\ \hline
          		$1.88\times 10^{4}-6.60\times 10^{4}$ &  $-2.28\pm 0.08$ & $7$ & $0.86$ \\ \hline
          		$8.17\times 10^{4}-1.88 \times 10^{6}$ & $-3.13\pm 0.01$ & $18$ & $3.44$ \\ \hline
          		
 \end{tabular}
 \caption{The values of the parameters obtained by fitting  $C^E_{\ell}$ of Cas A SNR}
 \label{tab:CasAfit}
 \end{table}
 
 \begin{table}
 \centering
 \begin{tabular}{c|c|c|c|}
          			\hline
          			$\ell$ range & power index $\beta$ & No. of points $N$ & $\frac{\chi^{2}}{(N-2)}$\\ \hline
          			$2.20\times 10^{4}-3.14 \times 10^{5}$ &  $-3.23\pm 0.01$ & $12$ & $1.56$ \\ \hline
          			$3.14\times 10^{5}-9.42\times 10^{5}$ & $-3.39\pm 0.01$ & $5$ & $1.03$ \\ \hline
          			
 \end{tabular}
 \caption{The values of the parameters obtained by fitting  $C^E_{\ell}$ of Crab SNR}
 \label{tab:Crabfit}
 \end{table}

\section{Results}
\label{sec:results}
Figure \ref{fig:AIPSkeplerCC} shows that the radio emission of the Kepler SNR has an incomplete shell and is slightly more elongated along the east-west direction. The asymmetry of north-south brightness distribution was explained by a bow shock model (\citet{Bandiera87},\cite{Borkowski1992}) where it was proposed that a bow shock structure in the North originated from the motion of a mass losing system through the ISM prior to the supernova. Recently, \cite{Zhang18} showed that the stellar winds of  Kepler SNR shape up the remnant as it strongly interacts with the dense circumstellar medium (CSM). Although the morphology and dynamics of Kepler SNR has been studied with observations \citep{DeLaney02} and numerical simulations \citep{ToledoRoy14}, the statistical properties of the small-scale structures associated with the remnant as quantified by the power spectrum (or correlation function) of observed intensity fluctuations has not been investigated in earlier works.

\begin{figure*}
 \begin{center}
   \psfrag{CK}[top]{ $\ell(\ell+1)\,C^E_\ell/2\pi \quad {\rm K}^2$}  \psfrag{ l}[80][b]{$\ell$} \psfrag{l}[80][b]{$\ell$} \psfrag{m}{$\ell_m$}
   \psfrag{10}{}
  \psfrag{2}[bottom]{$10^2$} \psfrag{3}[bottom]{$10^3$} \psfrag{4}[bottom]{$10^4$} \psfrag{5}[bottom]{$10^5$} \psfrag{6}[bottom]{$10^6$} \psfrag{7}[bottom]{$10^7$}
  \psfrag{3}[right]{$10^3$} \psfrag{2}[right]{$10^2$} \psfrag{1}[right]{$10^1$} \psfrag{0}[right]{$10^0$} \psfrag{-1}[right]{$10^{-1}$} \psfrag{-2}[right]{$10^{-2}$} \psfrag{-3}{} \psfrag{-4}[right]{$10^{-4}$} \psfrag{-5}{} \psfrag{-6}[right]{$10^{-6}$} \psfrag{-7}{} \psfrag{-8}[right]{$10^{-8}$} \psfrag{-9}{} \psfrag{-10}[right]{$10^{-10}$}\psfrag{-11}{}\psfrag{-12}[right]{$10^{-12}$}
 \includegraphics[scale=0.30,angle=-90]{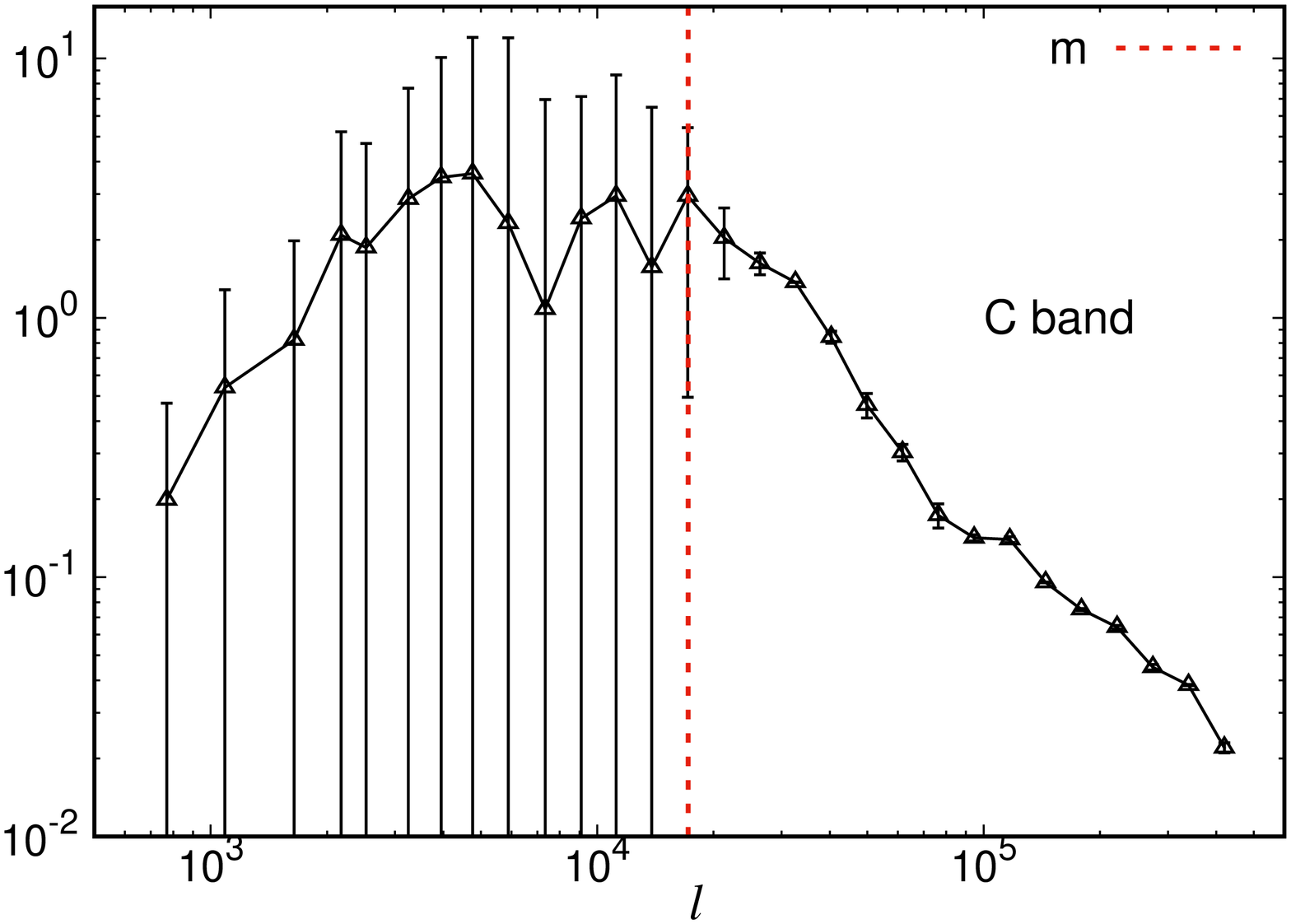}\hspace{0.5cm}\includegraphics[scale=0.30,angle=-90]{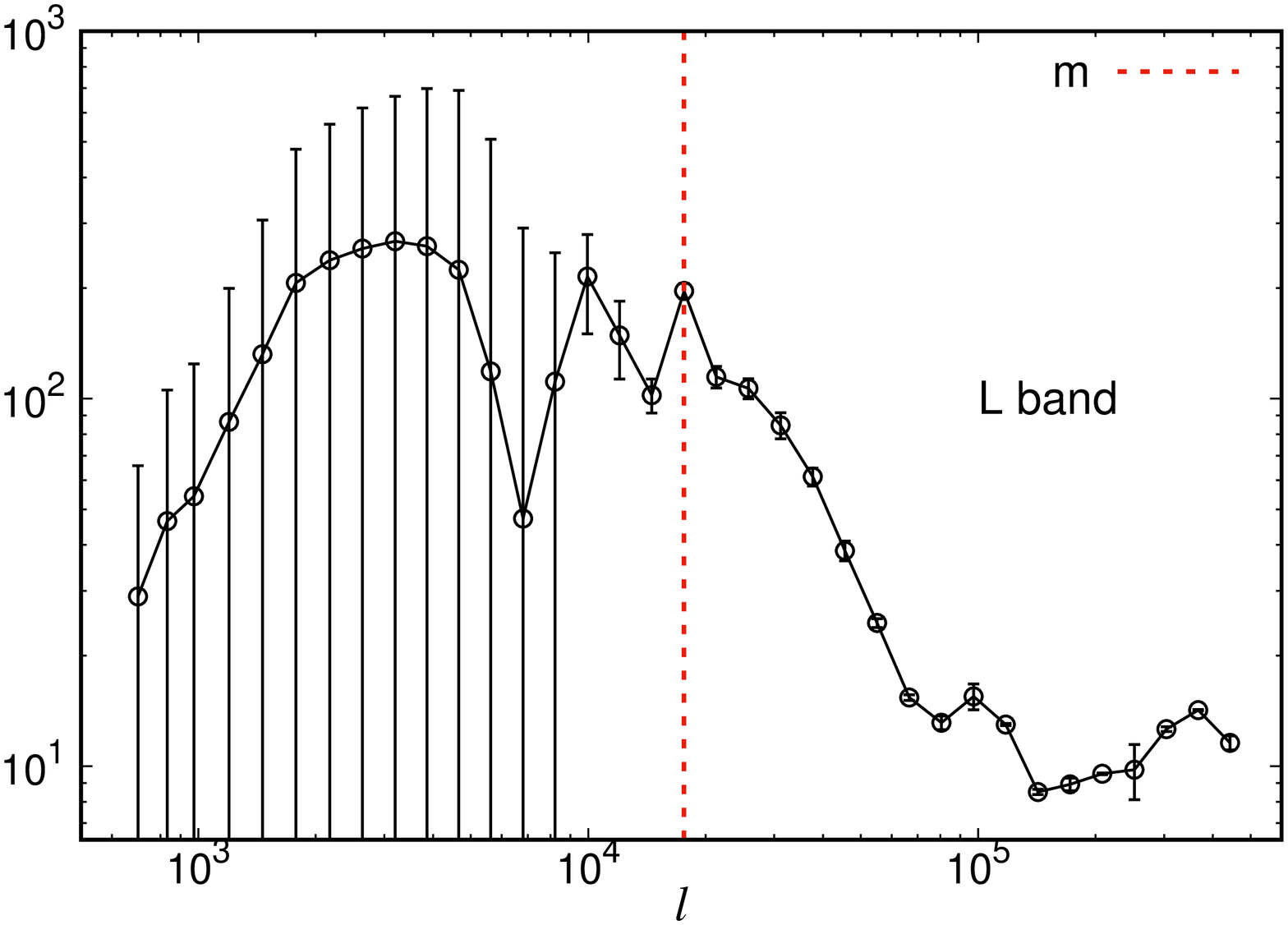}
 \caption{Estimated scaled  angular power spectrum    $ \ell (\ell+1)\, C^E_{\ell}/(2 \pi) $  of Kepler SNR as a function of angular multipole $\ell$. The $\pm 1\sigma$ error bars are derived by simulating the observations in A,B,C configurations of VLA C and L band.  }
  \label{fig:LCKepler}
 \end{center}
 \end{figure*}

\begin{figure}
\psfrag{ l}{} \psfrag{l}[bottom]{$\ell$} \psfrag{l1}{$\ell_1$} \psfrag{l2}{$\ell_2$}
\psfrag{C K}[top]{$\ell(\ell+1)\,C^E_\ell/2\pi$\quad [units]} \psfrag{10}{}
\psfrag{2}[bottom]{$10^2$} \psfrag{3}[bottom]{$10^3$} \psfrag{4}[bottom]{$10^4$} \psfrag{5}[bottom]{$10^5$} \psfrag{6}[bottom]{$10^6$} \psfrag{7}[bottom]{$10^7$}\psfrag{0}[right]{$10^0$} \psfrag{-1}[right]{$10^{-1}$}\psfrag{1}[right]{$10^1$}\psfrag{2}[right]{$10^2$}\psfrag{3}[right]{$10^3$}
\psfrag{-2}{} \psfrag{-3}[right]{$10^{-3}$} \psfrag{-4}{}\psfrag{-5}[right]{$10^{-5}$} \psfrag{-6}{} \psfrag{-7}[right]{$10^{-7}$} \psfrag{-8}{} \psfrag{-9}[right]{$10^{-9}$} \psfrag{-10}{}\psfrag{-11}[right]{$10^{-11}$}\psfrag{-12}{} \psfrag{-13}[right]{$10^{-13}$}\psfrag{-14}{}
\hspace{0.5cm}\includegraphics[scale=0.28,angle=-90]{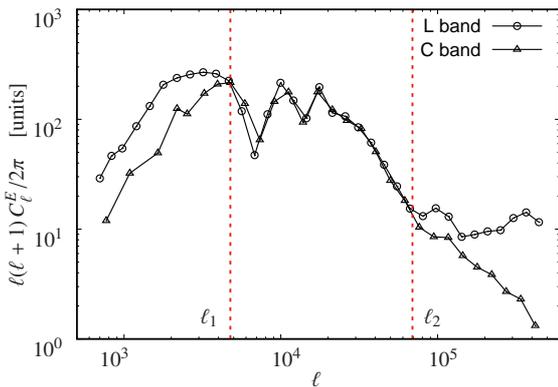}
\caption{Estimated scaled  angular power spectrum    $ \ell (\ell+1)\, C^E_{\ell}/(2 \pi) $ of Kepler SNR for C and L bands in arbitary units, without the $1\sigma$ error bars.}
\label{fig:KeplerC&L}
\end{figure}

Figure \ref{fig:LCKepler} presents the binned angular power spectrum $C^E_{\ell}$  estimated from the C band and L  band calibrated visibility data of the Kepler SNR (Table~\ref{tab:Keplerdata}. The  $\ell$ range  shown in the figure  corresponds to the entire baseline distribution analyzed here. For both C and L bands  at large $\ell$ we find that   $C^E_{\ell}$ falls steeply with increasing $\ell$  whereas this flattens out at small $\ell$ ($\leq \ell_m$)  where $\ell_m=1.7 \times 10^{4}$. The corresponding angular scale $\pi/\ell_m \approx 0.63^{'}$ is approximately $1/3$ the radius  of $(1.75^{'})$  of Kepler SNR estimated from Figure \ref{fig:AIPSkeplerCC}. The angular extent of the Kepler SNR is imprinted in the  angular profile function $\R$. As discussed earlier, the flattening seen at $\ell < \ell_m$ is an outcome of the  convolution (eq. \ref{eq:snr.b}) with the angular profile function $\R$. We notice that the error bars are rather large for $\ell < \ell_m$. The errors at small $\ell$ are sample variance dominated, and these are large as we do not have many independent estimates of the angular power spectrum at angular scales which are comparable or larger than the angular extent of the SNR. In addition to the steep decline in $C^E_{\ell}$ at $\ell>\ell_m$ and a flattening at $\ell < \ell_m$, we also notice two kink like features in both the C band and L band results around $\ell \sim 10^4$.  In order to check whether these are genuine astrophysical features associated with the Kepler SNR or if they are artifacts introduced by the baseline distribution, the estimator or some other effect we have superimposed the results from the two different bands after suitably scaling $C^E_{\ell}$ (Figure \ref{fig:KeplerC&L}). We find that the kink like features seen in the  C and L bands match with respect to both the $\ell$ position  as well as the relative amplitude. We also see that the C and L band results are in close agreement over a broad  $\ell$ range $\ell_1= 4.76 \times 10^3$ to $\ell_2= 6.91 \times 10^4$. The close match between the C and L band results further reinforces the idea that the estimated angular power spectrum $C^E_{\ell}$ reflects  genuine astrophysical features pertaining to the Kepler SNR. Considering the kinks in  $C^E_{\ell}$, the fact that the corresponding $\ell$ values  are smaller than $\ell_m$ indicates that  these are possibly related to morphological features in the overall profile of the Kepler SNR. In the subsequent parts of this paper we primarily focus on the range $\ell > \ell_m$ where $C^E_{\ell}$  fall steeply with increasing $\ell$. The C and L band results show the same (nearly) scale invariant behaviour in the range $\ell_m < \ell <\ell_2$. The scale invariant behaviour extends beyond $\ell_2$ for the C band  whereas the L band result flattens out at $\ell > \ell_2$. Later in this paper we discuss possible causes for this discrepancy between the C and L band results at large $\ell$.

\begin{figure*}
  
             	\begin{center}
             	  \psfrag{ l}{} \psfrag{l}[bottom]{$\ell$} 
             	 \psfrag{C K}[top]{$\ell(\ell+1)\,C^E_\ell/2\pi$ \quad ${\rm K}^2$}  \psfrag{10}{}
             	 \psfrag{2}[bottom]{$10^2$} \psfrag{3}[bottom]{$10^3$} \psfrag{4}[bottom]{$10^4$} \psfrag{5}[bottom]{$10^5$} \psfrag{6}[bottom]{$10^6$} \psfrag{7}[bottom]{$10^7$}
              \psfrag{1}[right]{$10^1$}  \psfrag{2}[right]{$10^2$}  \psfrag{3}[right]{$10^3$} 	 \psfrag{0}[right]{$10^0$} \psfrag{-1}[right]{$10^{-1}$} \psfrag{-2}[right]{$10^{-2}$} \psfrag{-3}[right]{$10^{-3}$} \psfrag{-4}[right]{$10^{-4}$} \psfrag{-5}[right]{$10^{-5}$} \psfrag{-6}{} \psfrag{-7}[right]{$10^{-7}$} \psfrag{-8}{} \psfrag{-9}[right]{$10^{-9}$} \psfrag{-10}{}\psfrag{-11}[right]{$10^{-11}$}\psfrag{-12}{}
             	 \psfrag{-13}[right]{$10^{-13}$}\psfrag{-14}{}
             		\includegraphics[scale=0.30,angle=-90]{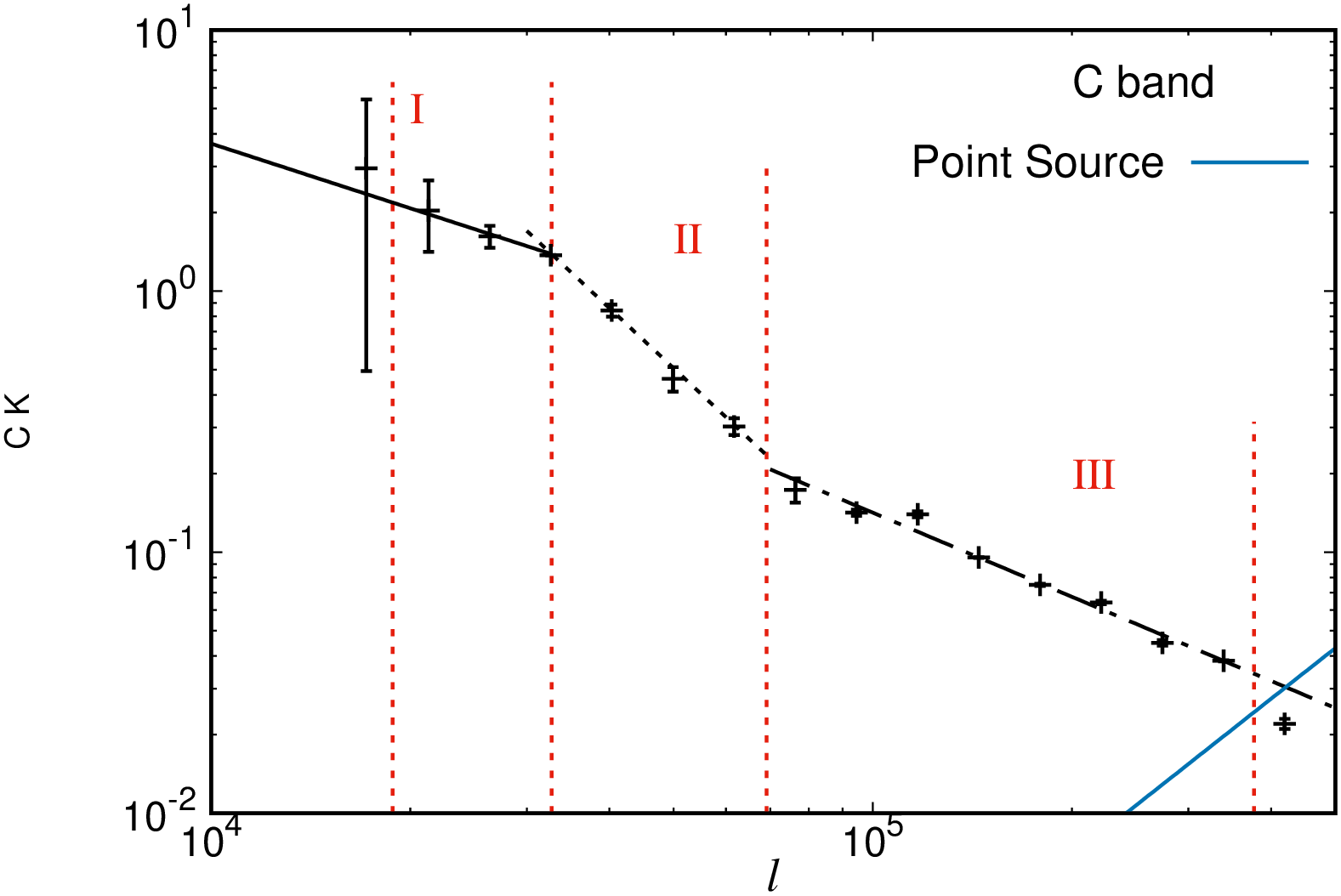}\hspace{0.5cm}	\includegraphics[scale=0.30,angle=-90]{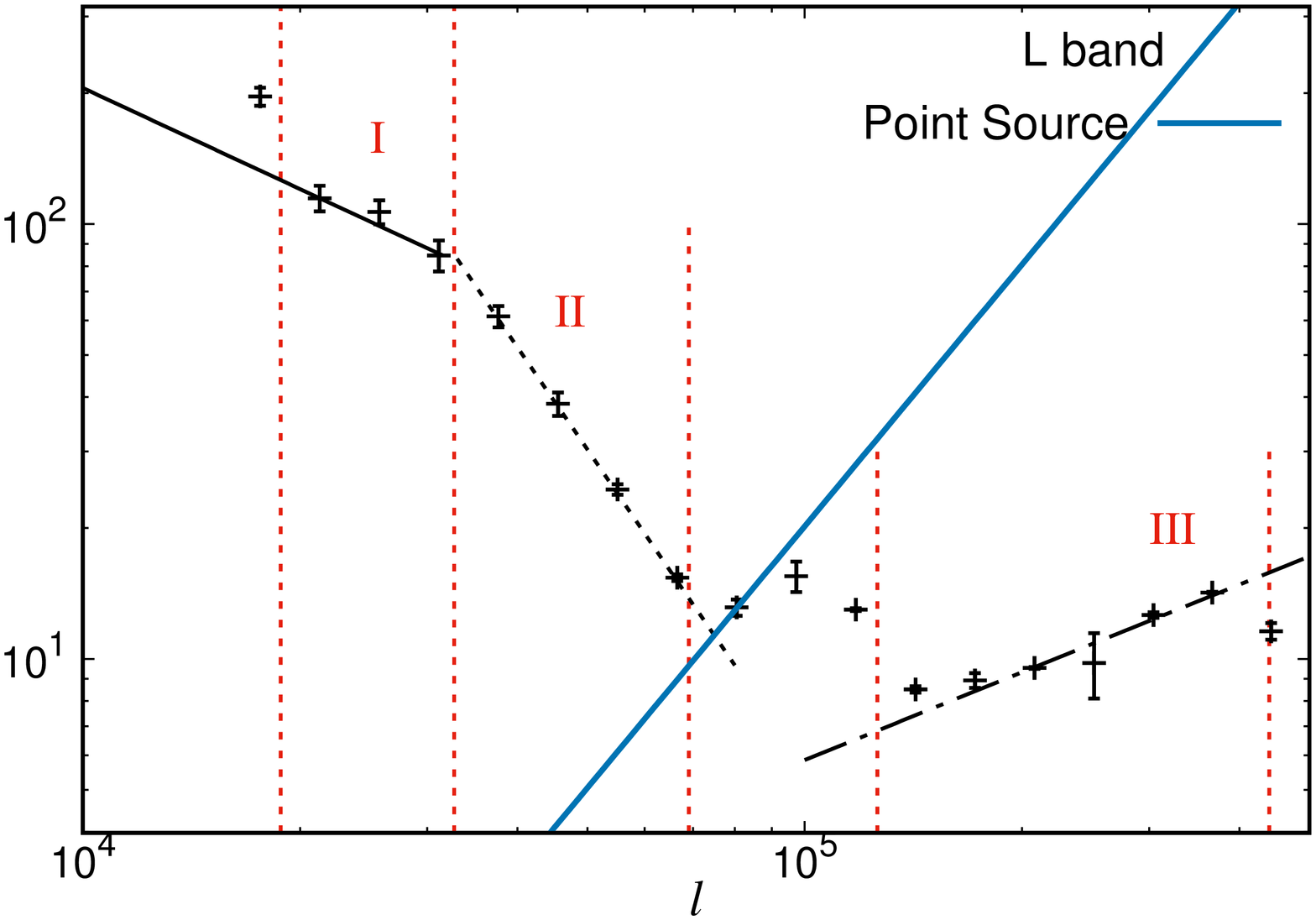}
             		\caption{Estimated scaled  angular power spectrum    $ \ell (\ell+1)\, C^E_{\ell}/(2 \pi) $ of Kepler SNR as a function of angular multipole $\ell$. The $\pm 1\sigma$ error bars are derived by simulating the observations of A,B,C configurations of VLA C and L bands. The convolution dominated region for $\ell<1.7 \times10^4$ has not been shown here. The best fit lines $C^E_{\ell}\propto \ell^{\,\beta}$ are plotted in black solid line, dot line and dot-dash line for $\ell$ ranges I,II and III respectively. The corresponding fit values of the  power-law index $\beta$ for the three $\ell$ ranges are listed in Table $\ref{tab:KepCfit}$ and Table $\ref{tab:KepLfit}$ for C and L band respectively.}
             		\label{fig:LCKeplerfit}
             \end{center}
             \end{figure*}

The subsequent analysis is restricted to the range $\ell > \ell_m$ where the shape of $C^E_{\ell}$  is not  affected by  the overall angular size of the remnant. For both the C and L bands we notice (Figure \ref{fig:LCKeplerfit}) that it is possible to identify three distinct $\ell$ ranges (respectively denoted as I, II and III) where $C^E_{\ell}$ shows a different behaviour in each of these $\ell$ ranges. We have separately analyzed $C^E_{\ell}$ for each $\ell$ range and observational band,  and carried out $\chi^2$ minimization to determine the best fit power-law of the form $C=A\ell^{\beta}$. Range I  spans $1.88\times10^4 < \ell \leq 3.27\times10^4$, and we have $\beta=-2.82\pm 0.06$ and  $\beta=-2.77\pm 0.26$ in the C and L bands respectively. Range II spans the range  $3.27\times10^4 < \ell \leq 6.91\times10^4$ and we have $\beta=-4.37\pm 0.06$ and $\beta=-4.43\pm 0.01$ in the C and  L bands respectively. In both of these $\ell$ ranges the results from the C band are consistent with those in the L band.  We find that $C^E_{\ell}$ shows a steepening from range I to range II  with $\beta$ increasing from  $\approx -2.8$ to  $\approx -4.4$ across the boundary at $\ell =3.27 \times 10^4$.  As noted earlier, the results from the C and L bands are different at large $\ell$ and as a consequence the $\ell$ boundaries of range III are different for each band. Considering the C band, we find that range III extends across a relatively large $\ell$ range spanning $ 6.91\times10^4 < \ell \leq 3.77\times 10^5$  where we  obtain a best-fit   power law with $\beta=-3.07\pm 0.02$. In the L band we find that range III extends across a relatively smaller $\ell$ range spanning  $1.26\times10^5 \leq \ell \leq 4.40\times10^5$ where we  obtain a best-fit   power law with $\beta=-1.33\pm 0.04$. The results of power-law fitting for ranges I, II and III  are summarized in Tables \ref{tab:KepCfit} and  \ref{tab:KepLfit} for the C and L bands respectively. We note that the reduced $\chi^{2}$ is particularly small for a few cases in ranges I and II, we interpret this as indicating that the errors (possibly sample variance) have been over-estimated at low $\ell$. In contrast we have quite large values of the   reduced $\chi^{2}$ in range III, we interpret this as indicating that the errors at large $\ell$ (mainly system noise) have been underestimated. In addition to this, there possibly are small deviations from the power law which contribute to the large  reduced $\chi^{2}$ at large $\ell$. Finally, we have suitably scaled $C^E_{\ell}$ estimated in the C and L bands and combined the  values. Considering the $\ell$ ranges I and II, a power law was fitted to the combined results for which the best fit parameters are presented in Table~\ref{tab:KepL&Cfit}. We find that the results from the combined analysis of the two bands yields $\beta=-2.84 \pm 0.07$  and $\beta=-4.39\pm 0.04$ for ranges I and II respectively. These results are consistent with those obtained individually from the C and L bands 
(Table \ref{tab:KepCfit} and  \ref{tab:KepLfit}).

We notice that in range III the  power law index has two different values $\beta=-3.07\pm 0.02$ and $\beta=-1.33\pm 0.04$ in the C and L bands respectively. Here the source Kepler SNR, with a flux density of $19$Jy at $1$GHz \citep{Green2014}, is much fainter than the  Cas A SNR ($2400$Jy at $1$GHz) and the Crab SNR ($960$Jy at $1$GHz) which have been  previously analyzed in \citet{roy09}. 
Therefore, it is possible that the angular power spectrum at small angular scales (large $\ell$) has significant contribution from other sources  like the extragalactic point sources  and the DGSE, collectively referred to as foregrounds. These foregrounds are frequency dependent and  its effects are more pronounced in the low frequency band i.e. L band as compared to C band. We estimate the expected contribution of the individual foregrounds from models which have been  developed in the context of $21$cm cosmology. The details are given in the Appendix \ref{appendix:a} and the results are shown in the Figure \ref{fig:LCKeplerfit}. For the C band, the expected  foreground contribution  is smaller than  the measured $C^E_{\ell}$ for nearly the entire $\ell$ range. However, for the L band the expected  foreground contribution  is comparable to  the measured $C^E_{\ell}$ for range III while the measured signal exceeds the predicted foregrounds in ranges I and II. We conclude that range III of L band is possibly contaminated by the foregrounds and we exclude this from the subsequent discussion. The rest of the  $C^E_{\ell}$ measurements  {\it i.e.} ranges I,II and III for C band and ranges I and II for L band, are not significantly affected by foregrounds and the measurements here pertain to the Kepler SNR.

            \begin{table}
                      	\centering
                      	\begin{tabular}{p{0.5cm}p{1.2cm}p{1.2cm}p{1.6cm}p{1cm}p{0.5cm}}
                      		\hline
                      		 Range  & $\ell_{min}$ & $\ell_{max}$ & power index $\beta$ & No. of points $N$ & $\frac{\chi^{2}}{(N-2)}$\\ \hline
                      		I & $1.88\times 10^{4}$ & $3.27\times 10^{4}$ &  $-2.82\pm 0.06$ & $3$ & $0.02$ \\ \hline
                      	 II & $3.25\times 10^{4}$ & $6.91\times 10^{4}$ & $-4.37\pm 0.06$ & $4$ & $0.35$ \\ \hline
                      	III & $6.91\times 10^{4}$ & $3.77\times 10^{5}$ & $-3.07\pm 0.02$ & $8$ & $9.89$ \\ \hline
                      	\end{tabular}
                      		\caption{The values of the parameters obtained by fitting $C^E_{\ell}$ of Kepler SNR in C band}
                      		\label{tab:KepCfit}
                      		\centering
                      		\begin{tabular}{p{0.5cm}p{1.2cm}p{1.2cm}p{1.6cm}p{1cm}p{0.5cm}}
                      			\hline
                      		 Range & $\ell_{min}$ & $\ell_{max}$ &power index $\beta$ & No. of points $N$ & $\frac{\chi^{2}}{(N-2)}$\\ \hline
                      		I & $1.88\times 10^{4}$ &  $3.27\times 10^{4}$ & $-2.77\pm 0.26$ & $3$ & $0.90$ \\ \hline
                      		II & $3.27\times 10^{4}$ & $6.91\times 10^{4}$ & $-4.43\pm 0.01$ & $4$ & $0.02$ \\ \hline
                      		III & $1.26\times 10^{5}$ & $4.40\times 10^{5}$ & $-1.33\pm 0.04$ & $6$ & $16.30$ \\ \hline
                      		\end{tabular}
                      	\caption{The values of the parameters obtained by fitting $C^E_{\ell}$ of Kepler SNR in L band}
                      	\label{tab:KepLfit}
                      	\centering
                      	  \begin{tabular}{p{0.5cm}p{1.2cm}p{1.2cm}p{1.6cm}p{1cm}p{0.5cm}}
                      	      \hline
                      Range & $\ell_{min}$ & $\ell_{max}$ & power index $\beta$ & No. of points $N$ & $\frac{\chi^{2}}{(N-2)}$\\ \hline
                     	I  & $1.88\times 10^{4}$ &  $3.27\times 10^{4}$ & $-2.84\pm 0.07$ & $6$ & $0.27$ \\ \hline
                       II & $3.27 \times 10^{4}$ & $6.91 \times 10^{4}$ & $-4.39\pm 0.04$ & $7$ & $0.29$ \\ \hline
                      	    
                      	                      		\end{tabular}
                      	    \caption{The values of the parameters obtained by fitting combined $C^E_{\ell}$ of Kepler SNR in L and C bands}
                      	    \label{tab:KepL&Cfit}
                      \end{table}

\section{Discussion and Conclusion}
\label{sec:conclusion}
At angular scales $\ell > \ell_m$, we observe  that $C^E_{\ell}$ for Kepler SNR (Figure \ref{fig:LCKeplerfit}) shows three different  power laws  (Table \ref{tab:KepCfit} and \ref{tab:KepLfit}) over the three distinct  $\ell$ ranges (I, II and III). Considering range I which has the smallest  $\ell$ values we find  the power law index $\beta=-2.8$ for both the bands. This value is comparable to the power law index $\beta=-8/3$ predicted for  two-dimensional in-compressible Kolmogorov turbulence \citep{kolmogorov1941}. This is also consistent with the findings of \cite{shimoda2018} who have analyzed  the intensity fluctuations of a $1.4$GHz VLA image of the Tycho SNR and found that  the two-point correlation function in the outer shells of the SNR shows a scaling close to  the Kolmogorov  $r^{2/3}$. This kind of power law scaling is also predicted for developed MHD turbulence \citep{GS95}.
Considering the Kepler SNR, we see that the power law becomes steeper in range II with $\beta=-4.4$.    
\cite{roy09} have found a  steepening of the power law index from $-2.2$ at $\ell \le 6.28\times10^4$ to $-3.2$ at $\ell > 6.91\times10^4$ for Cas A SNR. It is well accepted that Cas A is a shell type SNR \citep{Reed95} and the break in the power law was interpreted as a transition from 2D to 3D turbulence at the scale corresponding to the shell thickness. Interpreting the transition from range I to range II of Kepler SNR  along the same lines, we see that the boundary at $\ell=3.27 \times 10^4$  corresponds to a shell thickness of $0.48$pc which roughly matches with the value of $0.35$pc reported in \cite{matsui1984} and  \cite{matsui1988}. We would expect $\beta$ to change from $-8/3$ to $-11/3$ for Kolmogorov turbulence. However, the difficulty for Kepler SNR is that the observed steepening from $-2.8$ to $-4.4$  significantly exceeds the difference of $-1$ expected in a transition from 2D to 3D turbulence.  
 Considering range III, we see that the power law again becomes flatter with $\beta=-3.1$.  This is close to the power law index $\beta =-3.2$ found at large $\ell$ for both the Cas A and the Crab SNRs \citep{roy09}.  The equivalent energy spectrum $E(k)=k^2 P(k)$
for Kepler SNR is  $E(k)\propto k^{-1.1}$. One plausible reason for the slope in range III being less steep as compared to range II is that  viscous damping can be important in small-scale turbulence. \cite{Cho2002} have shown that the turbulent magnetic energy spectrum  is flattened due to viscous damping, however the relation between the magnetic power spectrum and  the intensity power spectrum reported here is not well understood. 
 
\begin{figure}
\psfrag{C K}[2]{$\ell(\ell+1)\,C_\ell/2\pi$  \quad ${\rm K}^{2}$} 
\psfrag{ l}{} \psfrag{l}[bottom]{$\ell$} \psfrag{l1}{$\ell_1$} \psfrag{l2}{$\ell_2$}
 \psfrag{10}{}
\psfrag{2}[bottom]{$10^2$} \psfrag{3}[bottom]{$10^3$} \psfrag{4}[bottom]{$10^4$} \psfrag{5}[bottom]{$10^5$} \psfrag{6}[bottom]{$10^6$} \psfrag{7}[bottom]{$10^7$}\psfrag{1}[right]{$10^1$}\psfrag{0}[right]{$10^0$} \psfrag{-1}[right]{$10^{-1}$}
\psfrag{-2}{} \psfrag{-3}[right]{$10^{-3}$} \psfrag{-4}{}\psfrag{-5}[right]{$10^{-5}$} \psfrag{-6}{} \psfrag{-7}[right]{$10^{-7}$} \psfrag{-8}{} \psfrag{-9}[right]{$10^{-9}$} \psfrag{-10}{}\psfrag{-11}[right]{$10^{-11}$}\psfrag{-12}{} \psfrag{-13}[right]{$10^{-13}$}\psfrag{-14}{}
\hspace{0.5cm}\includegraphics[scale=0.28,angle=-90]{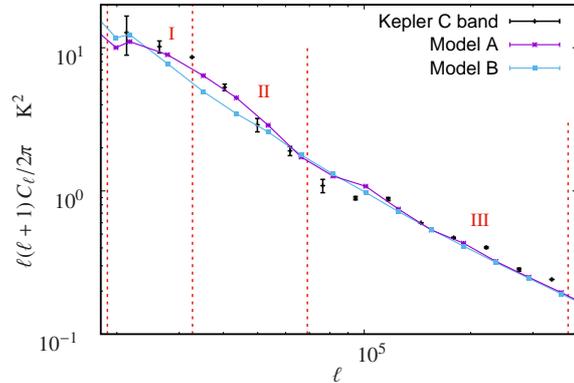}
\caption{Comparison of the model angular power spectrum $C_{\ell}$ with the estimated C band  angular power spectrum $C^E_{\ell}$ of the Kepler SNR. Models A and B shown here differ in the value of the inner radius of the shell which is  $1.2^{\arcmin}$ and $0.9^{\arcmin}$ respectively.}
\label{fig:sim_model_fit}
\end{figure}

To validate our above interpretations of $C^E_{\ell}$, we present systematic 3D simulations which were carried out to model the envisaged scenario. These simulations aid to understand the effect of the finite shell thickness and also investigate the effect of the line of sight integration on the measured signal. Since the entire observations and analysis are in terms of the angular scale $\theta$ and the angular multipoles $\ell$, the 3D simulations are in terms of ${\bf r}$ which is in angular units and its  Fourier conjugate ${\bf k}$ for which $\mid {\bf k}\mid = \ell$. These can be converted to  physical units using  the distance to the SNR. We simulate the brightness temperature fluctuations $\delta T({\bf r})$ inside a 3D cube of size $[3.072^{\arcmin^3}]$ with $[1024]^3$ grid points and spacing $0.003^{\arcmin}$. We ascribe a shell geometry to the remnant to model range I and II. For this, we generate  fluctuations $\delta T({\bf r})$ considering a 3D power spectrum $P(k)=A_{1}k^{-4.4}$ which corresponds to range II. To mimic the Kepler SNR's shell, we introduce a spherical shell of outer radius $1.5^{\arcmin}$ (which is approximately the observed size of the SNR) and shell  thickness $0.3^{\arcmin}$ which approximately corresponds to $\ell=3.27 \times 10^4$  where we have the boundary between range I and II.  The off-shell brightness temperature fluctuations were erased. To model range III of C band, we generate  brightness temperature fluctuations corresponding to a 3D power spectrum $P(k)=A_{2}k^{-3.1}$. These fluctuations fill the core which is a sphere of radius $1.2^{\arcmin}$ enclosed within the inner radius  of the shell.  The shell and the core are combined and the resulting 3D cube is projected on to a 2D plane which corresponds to the plane of the sky. The amplitudes $A_{1}$ and $A_{2}$ are set to match the data $C^E_{\ell}$ of C band (left panel of Figure \ref{fig:LCKeplerfit}). We estimate the  angular power spectrum $C_{\ell}$ from the projected 2D brightness temperature fluctuations. We have used $1,000$ statistically independent realizations of the simulations to obtain the mean  $C_{\ell}$ shown as Model A in Figure \ref{fig:sim_model_fit}.  We see that the  simulated  $C_{\ell}$ is able to reproduce the features visible in $C^E_{\ell}$. For comparison we have also considered Model B where the shell thickness is doubled to $0.6^{\arcmin}$. We find that  Model B fails to match  $C^E_{\ell}$ in the vicinity of the transition from range I to range II. This indicates that the shell thickness is reflected in the $\ell$ position of the transition from range I to II. These simulations validate our geometrical picture  where we interpret  range I as 2D turbulence at angular scales larger than the shell thickness, range II as 3D turbulence within the shell and range III as 3D turbulence within the core. The line of sight  averaging does not affect the slope of the estimated power spectrum in ranges II and III. We finally note that our model presented here is not unique, rather it presents a plausible geometrical picture of the SNR.

In this paper we have interpreted the intensity fluctuations of the Kepler SNR as arising from MHD turbulence. 
At large angular scales the slope $(\beta=-2.8)$ of the measured power spectrum is consistent with 2D Kolmogorov turbulence 
and also the angular two-point correlation of the Tycho SNR \citep{shimoda2018}. At small angular scales the slope $(\beta=-3.1)$ for the Kepler SNR  is consistent with earlier measurements for Cas A and Crab SNRs \citep{roy09}. However, for Kepler we have a third  intermediate $\ell$ range where the power spectrum falls steeply with $\beta=-4.4$. Such a steep intermediate range has not been observed in any of the three SNRs which have been analyzed earlier and this is possibly an outcome of the complex morphology of the Kepler SNR (Figure \ref{fig:AIPSkeplerCC}).  The relation between the slopes of the power spectrum  for different SNRs is expected to throw light on the nature of the underlying physical processes. 
We plan to carry out similar power spectrum analysis for other Galactic supernova remnants in future.

\section*{Acknowledgements}
 P. S. is thankful to Prasun Dutta for useful discussions. N.~R.~acknowledges support from the Infosys Foundation through the Infosys Young Investigator grant. We also thank the anonymous referee and the editor for their useful comments. 

\bibliographystyle{mnras}

\bibliography{bibreportext}

\begin{thebibliography}{}
\makeatletter
\relax
\def\mn@urlcharsother{\let\do\@makeother \do\$\do\&\do\#\do\^\do\_\do\%\do\~}
\def\mn@doi{\begingroup\mn@urlcharsother \@ifnextchar [ {\mn@doi@}
  {\mn@doi@[]}}
\def\mn@doi@[#1]#2{\def\@tempa{#1}\ifx\@tempa\@empty \href
  {http://dx.doi.org/#2} {doi:#2}\else \href {http://dx.doi.org/#2} {#1}\fi
  \endgroup}
\def\mn@eprint#1#2{\mn@eprint@#1:#2::\@nil}
\def\mn@eprint@arXiv#1{\href {http://arxiv.org/abs/#1} {{\tt arXiv:#1}}}
\def\mn@eprint@dblp#1{\href {http://dblp.uni-trier.de/rec/bibtex/#1.xml}
  {dblp:#1}}
\def\mn@eprint@#1:#2:#3:#4\@nil{\def\@tempa {#1}\def\@tempb {#2}\def\@tempc
  {#3}\ifx \@tempc \@empty \let \@tempc \@tempb \let \@tempb \@tempa \fi \ifx
  \@tempb \@empty \def\@tempb {arXiv}\fi \@ifundefined
  {mn@eprint@\@tempb}{\@tempb:\@tempc}{\expandafter \expandafter \csname
  mn@eprint@\@tempb\endcsname \expandafter{\@tempc}}}

\bibitem[\protect\citeauthoryear{{Bandiera}}{{Bandiera}}{1987}]{Bandiera87}
{Bandiera} R.,  1987, \mn@doi [\apj] {10.1086/165505}, \href
  {https://ui.adsabs.harvard.edu/\#abs/1987ApJ...319..885B} {319, 885}

\bibitem[\protect\citeauthoryear{{Begum}, {Chengalur}  \& {Bhardwaj}}{{Begum}
  et~al.}{2006}]{begum2006}
{Begum} A.,  {Chengalur} J.~N.,   {Bhardwaj} S.,  2006, \mn@doi [\mnras]
  {10.1111/j.1745-3933.2006.00220.x}, \href
  {http://adsabs.harvard.edu/abs/2006MNRAS.372L..33B} {372, L33}

\bibitem[\protect\citeauthoryear{{Bondi} et~al.,}{{Bondi}
  et~al.}{2003}]{Bondi2003}
{Bondi} M.,  et~al., 2003, \mn@doi [\aap] {10.1051/0004-6361:20030382}, \href
  {https://ui.adsabs.harvard.edu/#abs/2003A&A...403..857B} {403, 857}

\bibitem[\protect\citeauthoryear{{Borkowski}, {Blondin}  \&
  {Sarazin}}{{Borkowski} et~al.}{1992}]{Borkowski1992}
{Borkowski} K.~J.,  {Blondin} J.~M.,   {Sarazin} C.~L.,  1992, \mn@doi [\apj]
  {10.1086/171989}, \href {http://adsabs.harvard.edu/abs/1992ApJ...400..222B}
  {400, 222}

\bibitem[\protect\citeauthoryear{Chakraborty et~al.,}{Chakraborty
  et~al.}{2019}]{Chakraborty2019}
Chakraborty A.,  et~al., 2019, \mn@doi [\mnras] {10.1093/mnras/stz1580}, \href
  {https://doi.org/10.1093/mnras/stz1580} {487, 4102}

\bibitem[\protect\citeauthoryear{{Chandrasekhar} \&
  {M{\"u}nch}}{{Chandrasekhar} \& {M{\"u}nch}}{1952}]{CM_V1952}
{Chandrasekhar} S.,  {M{\"u}nch} G.,  1952, \mn@doi [\apj] {10.1086/145518},
  \href {https://ui.adsabs.harvard.edu/abs/1952ApJ...115..103C} {115, 103}

\bibitem[\protect\citeauthoryear{Chengalur, Gupta  \& Dwarakanath}{Chengalur
  et~al.}{2003}]{gmrt.book}
Chengalur J.,  Gupta Y.,   Dwarakanath K.~S.,  2003, Low frequency Radio
  Astronomy, NCRA-TIFR

\bibitem[\protect\citeauthoryear{{Chevalier}}{{Chevalier}}{1977}]{Chevalier1977}
{Chevalier} R.~A.,  1977, \mn@doi [\araa]
  {10.1146/annurev.aa.15.090177.001135}, \href
  {http://adsabs.harvard.edu/abs/1977ARA%26A..15..175C} {15, 175}

\bibitem[\protect\citeauthoryear{Cho, Lazarian  \& Vishniac}{Cho
  et~al.}{2002}]{Cho2002}
Cho J.,  Lazarian A.,   Vishniac E.~T.,  2002, \mn@doi [\apj] {10.1086/339453},
  \href {https://doi.org/10.1086%2F339453} {566, L49}

\bibitem[\protect\citeauthoryear{{Choudhuri}, {Bharadwaj}, {Ghosh}  \&
  {Ali}}{{Choudhuri} et~al.}{2014}]{Choudhuri2014}
{Choudhuri} S.,  {Bharadwaj} S.,  {Ghosh} A.,   {Ali} S.~S.,  2014, \mn@doi
  [\mnras] {10.1093/mnras/stu2027}, \href
  {http://adsabs.harvard.edu/abs/2014MNRAS.445.4351C} {445, 4351}

\bibitem[\protect\citeauthoryear{{Choudhuri}, {Bharadwaj}, {Roy}, {Ghosh}  \&
  {Ali}}{{Choudhuri} et~al.}{2016a}]{Choudhuri2016a}
{Choudhuri} S.,  {Bharadwaj} S.,  {Roy} N.,  {Ghosh} A.,   {Ali} S.~S.,  2016a,
  \mn@doi [\mnras] {10.1093/mnras/stw607}, \href
  {https://ui.adsabs.harvard.edu/abs/2016MNRAS.459..151C} {459, 151}

\bibitem[\protect\citeauthoryear{{Choudhuri}, {Bharadwaj}, {Chatterjee}, {Ali},
  {Roy}  \& {Ghosh}}{{Choudhuri} et~al.}{2016b}]{Choudhuri2016b}
{Choudhuri} S.,  {Bharadwaj} S.,  {Chatterjee} S.,  {Ali} S.~S.,  {Roy} N.,
  {Ghosh} A.,  2016b, \mn@doi [\mnras] {10.1093/mnras/stw2254}, \href
  {http://adsabs.harvard.edu/abs/2016MNRAS.463.4093C} {463, 4093}

\bibitem[\protect\citeauthoryear{{Choudhuri}, {Roy}, {Bharadwaj}, {Saiyad Ali},
  {Ghosh}  \& {Dutta}}{{Choudhuri} et~al.}{2017}]{2017NewA...57...94C}
{Choudhuri} S.,  {Roy} N.,  {Bharadwaj} S.,  {Saiyad Ali} S.,  {Ghosh} A.,
  {Dutta} P.,  2017, \mn@doi [\na] {10.1016/j.newast.2017.06.010}, \href
  {http://adsabs.harvard.edu/abs/2017NewA...57...94C} {57, 94}

\bibitem[\protect\citeauthoryear{{Ciliegi}, {Zamorani}, {Hasinger}, {Lehmann},
  {Szokoly}  \& {Wilson}}{{Ciliegi} et~al.}{2003}]{Ciliegi2003}
{Ciliegi} P.,  {Zamorani} G.,  {Hasinger} G.,  {Lehmann} I.,  {Szokoly} G.,
  {Wilson} G.,  2003, \mn@doi [\aap] {10.1051/0004-6361:20021721}, \href
  {https://ui.adsabs.harvard.edu/#abs/2003A&A...398..901C} {398, 901}

\bibitem[\protect\citeauthoryear{DeLaney, Koralesky, Rudnick  \&
  Dickel}{DeLaney et~al.}{2002}]{DeLaney02}
DeLaney T.,  Koralesky B.,  Rudnick L.,   Dickel J.~R.,  2002, \apj, \href
  {http://stacks.iop.org/0004-637X/580/i=2/a=914} {580, 914}

\bibitem[\protect\citeauthoryear{{Dickel}, {Sault}, {Arendt}, {Matsui}  \&
  {Korista}}{{Dickel} et~al.}{1988}]{matsui1988}
{Dickel} J.~R.,  {Sault} R.,  {Arendt} R.~G.,  {Matsui} Y.,   {Korista} K.~T.,
  1988, \mn@doi [\apj] {10.1086/166469}, \href
  {http://adsabs.harvard.edu/abs/1988ApJ...330..254D} {330, 254}

\bibitem[\protect\citeauthoryear{{Dutta}, {Begum}, {Bharadwaj}  \&
  {Chengalur}}{{Dutta} et~al.}{2008}]{Dutta2008}
{Dutta} P.,  {Begum} A.,  {Bharadwaj} S.,   {Chengalur} J.~N.,  2008, \mn@doi
  [\mnras] {10.1111/j.1745-3933.2007.00417.x}, \href
  {https://ui.adsabs.harvard.edu/abs/2008MNRAS.384L..34D} {384, L34}

\bibitem[\protect\citeauthoryear{{Dutta}, {Begum}, {Bharadwaj}  \&
  {Chengalur}}{{Dutta} et~al.}{2009}]{dutta2009}
{Dutta} P.,  {Begum} A.,  {Bharadwaj} S.,   {Chengalur} J.~N.,  2009, \mn@doi
  [\mnras] {10.1111/j.1365-2966.2009.15105.x}, \href
  {http://adsabs.harvard.edu/abs/2009MNRAS.398..887D} {398, 887}

\bibitem[\protect\citeauthoryear{{Dutta}, {Begum}, {Bharadwaj}  \&
  {Chengalur}}{{Dutta} et~al.}{2013}]{Dutta2013}
{Dutta} P.,  {Begum} A.,  {Bharadwaj} S.,   {Chengalur} J.~N.,  2013, \mn@doi
  [\na] {10.1016/j.newast.2012.08.008}, \href
  {https://ui.adsabs.harvard.edu/abs/2013NewA...19...89D} {19, 89}

\bibitem[\protect\citeauthoryear{{Ghosh}, {Prasad}, {Bharadwaj}, {Ali}  \&
  {Chengalur}}{{Ghosh} et~al.}{2012}]{Ghosh2012}
{Ghosh} A.,  {Prasad} J.,  {Bharadwaj} S.,  {Ali} S.~S.,   {Chengalur} J.~N.,
  2012, \mn@doi [\mnras] {10.1111/j.1365-2966.2012.21889.x}, \href
  {http://adsabs.harvard.edu/abs/2012MNRAS.426.3295G} {426, 3295}

\bibitem[\protect\citeauthoryear{{Goldreich} \& {Sridhar}}{{Goldreich} \&
  {Sridhar}}{1995}]{GS95}
{Goldreich} P.,  {Sridhar} S.,  1995, \mn@doi [\apj] {10.1086/175121}, \href
  {http://adsabs.harvard.edu/abs/1995ApJ...438..763G} {438, 763}

\bibitem[\protect\citeauthoryear{{Green}}{{Green}}{2014}]{Green2014}
{Green} D.~A.,  2014, Bulletin of the Astronomical Society of India, \href
  {http://adsabs.harvard.edu/abs/2014BASI...42...47G} {42, 47}

\bibitem[\protect\citeauthoryear{{Gull}}{{Gull}}{1973}]{Gull1973a}
{Gull} S.~F.,  1973, \mn@doi [\mnras] {10.1093/mnras/161.1.47}, \href
  {http://adsabs.harvard.edu/abs/1973MNRAS.161...47G} {161, 47}

\bibitem[\protect\citeauthoryear{{Gull}}{{Gull}}{1975}]{Gull1975}
{Gull} S.~F.,  1975, \mn@doi [\mnras] {10.1093/mnras/171.2.237}, \href
  {http://adsabs.harvard.edu/abs/1975MNRAS.171..237G} {171, 237}

\bibitem[\protect\citeauthoryear{{Katsuda} et~al.,}{{Katsuda}
  et~al.}{2015}]{Katsuda2015}
{Katsuda} S.,  et~al., 2015, \mn@doi [\apj] {10.1088/0004-637X/808/1/49}, \href
  {http://adsabs.harvard.edu/abs/2015ApJ...808...49K} {808, 49}

\bibitem[\protect\citeauthoryear{Kolmogorov}{Kolmogorov}{1941}]{kolmogorov1941}
Kolmogorov A.~N.,  1941, Proc. Acad. Sci. USSR, \href
  {https://ci.nii.ac.jp/naid/10021168141/en/} {31, 538}

\bibitem[\protect\citeauthoryear{{Matsui}, {Long}, {Dickel}  \&
  {Greisen}}{{Matsui} et~al.}{1984}]{matsui1984}
{Matsui} Y.,  {Long} K.~S.,  {Dickel} J.~R.,   {Greisen} E.~W.,  1984, \mn@doi
  [\apj] {10.1086/162688}, \href
  {http://adsabs.harvard.edu/abs/1984ApJ...287..295M} {287, 295}

\bibitem[\protect\citeauthoryear{Monin \& Yaglom}{Monin \&
  Yaglom}{1971}]{monin1971statistical}
Monin A.,  Yaglom A.,  1971, Statistical fluid mechanics; mechanics of
  turbulence, Cambridge, Mass., MIT Press [c1971-75]

\bibitem[\protect\citeauthoryear{{Patnaude}, {Badenes}, {Park}  \&
  {Laming}}{{Patnaude} et~al.}{2012}]{Patnaude}
{Patnaude} D.~J.,  {Badenes} C.,  {Park} S.,   {Laming} J.~M.,  2012, \mn@doi
  [\apj] {10.1088/0004-637X/756/1/6}, \href
  {http://adsabs.harvard.edu/abs/2012ApJ...756....6P} {756, 6}

\bibitem[\protect\citeauthoryear{{Reed}, {Hester}, {Fabian}  \&
  {Winkler}}{{Reed} et~al.}{1995}]{Reed95}
{Reed} J.~E.,  {Hester} J.~J.,  {Fabian} A.~C.,   {Winkler} P.~F.,  1995,
  \mn@doi [\apj] {10.1086/175308}, \href
  {http://adsabs.harvard.edu/abs/1995ApJ...440..706R} {440, 706}

\bibitem[\protect\citeauthoryear{{Reynolds}, {Borkowski}, {Hwang}, {Hughes},
  {Badenes}, {Laming}  \& {Blondin}}{{Reynolds} et~al.}{2007}]{Reynolds2007}
{Reynolds} S.~P.,  {Borkowski} K.~J.,  {Hwang} U.,  {Hughes} J.~P.,  {Badenes}
  C.,  {Laming} J.~M.,   {Blondin} J.~M.,  2007, \mn@doi [\apjl]
  {10.1086/522830}, \href {http://adsabs.harvard.edu/abs/2007ApJ...668L.135R}
  {668, L135}

\bibitem[\protect\citeauthoryear{{Reynoso} \& {Goss}}{{Reynoso} \&
  {Goss}}{1999}]{ReyGoss99}
{Reynoso} E.~M.,  {Goss} W.~M.,  1999, \mn@doi [\aj] {10.1086/300990}, \href
  {http://adsabs.harvard.edu/abs/1999AJ....118..926R} {118, 926}

\bibitem[\protect\citeauthoryear{{Roy}, {Bharadwaj}, {Dutta}  \&
  {Chengalur}}{{Roy} et~al.}{2009}]{roy09}
{Roy} N.,  {Bharadwaj} S.,  {Dutta} P.,   {Chengalur} J.~N.,  2009, \mn@doi
  [\mnras] {10.1111/j.1745-3933.2008.00591.x}, \href
  {http://adsabs.harvard.edu/abs/2009MNRAS.393L..26R} {393, L26}

\bibitem[\protect\citeauthoryear{{Sedov}}{{Sedov}}{1946}]{sedov1946}
{Sedov} L.~I.,  1946, Journal of Applied Mathematics and Mechanics, 10, 241

\bibitem[\protect\citeauthoryear{{Shimoda}, {Akahori}, {Lazarian}, {Inoue}  \&
  {Fujita}}{{Shimoda} et~al.}{2018}]{shimoda2018}
{Shimoda} J.,  {Akahori} T.,  {Lazarian} A.,  {Inoue} T.,   {Fujita} Y.,  2018,
  \mn@doi [\mnras] {10.1093/mnras/sty2034}, \href
  {https://ui.adsabs.harvard.edu/\#abs/2018MNRAS.480.2200S} {480, 2200}

\bibitem[\protect\citeauthoryear{{Singal}, {Stawarz}, {Lawrence}  \&
  {Petrosian}}{{Singal} et~al.}{2010}]{Singal2010}
{Singal} J.,  {Stawarz} {\L}.,  {Lawrence} A.,   {Petrosian} V.,  2010, \mn@doi
  [\mnras] {10.1111/j.1365-2966.2010.17382.x}, \href
  {http://adsabs.harvard.edu/abs/2010MNRAS.409.1172S} {409, 1172}

\bibitem[\protect\citeauthoryear{Taylor}{Taylor}{1935}]{taylor1935statistical}
Taylor G.~I.,  1935, Proceedings of the Royal Society of London. Series
  A-Mathematical and Physical Sciences, 151, 465

\bibitem[\protect\citeauthoryear{Taylor}{Taylor}{1950}]{Taylor1949}
Taylor G.~I.,  1950, Proceedings of the Royal Society of London. Series A,
  Mathematical and Physical Sciences, 201, 159

\bibitem[\protect\citeauthoryear{{Thompson}, {Moran}  \& {Swenson}}{{Thompson}
  et~al.}{2017}]{isra.book}
{Thompson} A.~R.,  {Moran} J.~M.,   {Swenson} Jr. G.~W.,  2017, {Interferometry
  and Synthesis in Radio Astronomy, 3rd Edition},
  \mn@doi{10.1007/978-3-319-44431-4.
}

\bibitem[\protect\citeauthoryear{Toledo-Roy, Esquivel, Velázquez  \&
  Reynoso}{Toledo-Roy et~al.}{2014}]{ToledoRoy14}
Toledo-Roy J.~C.,  Esquivel A.,  Velázquez P.~F.,   Reynoso E.~M.,  2014,
  \mn@doi [\mnras] {10.1093/mnras/stu880}, \href
  {http://dx.doi.org/10.1093/mnras/stu880} {442, 229}

\bibitem[\protect\citeauthoryear{{Wilson}, {Rohlfs}  \& {
  Huttemeister}}{{Wilson} et~al.}{2013}]{tora.book}
{Wilson} T.~L.,  {Rohlfs} K.,   { Huttemeister} S.,  2013, {Tools of Radio
  Astronomy, (6th Edition), ~Springer-Verlag (Berlin)},
  \mn@doi{10.1007/978-3-319-44431-4.
}

\bibitem[\protect\citeauthoryear{{Zhang} \& {Chevalier}}{{Zhang} \&
  {Chevalier}}{2019}]{Zhang18}
{Zhang} D.,  {Chevalier} R.~A.,  2019, \mn@doi [\mnras]
  {10.1093/mnras/sty2769}, \href
  {https://ui.adsabs.harvard.edu/\#abs/2019MNRAS.482.1602Z} {482, 1602}

\bibitem[\protect\citeauthoryear{{de Oliveira-Costa}, {Tegmark}, {Gaensler},
  {Jonas}, {Landecker}  \& {Reich}}{{de Oliveira-Costa}
  et~al.}{2008}]{deCosta2008}
{de Oliveira-Costa} A.,  {Tegmark} M.,  {Gaensler} B.~M.,  {Jonas} J.,
  {Landecker} T.~L.,   {Reich} P.,  2008, \mn@doi [\mnras]
  {10.1111/j.1365-2966.2008.13376.x}, \href
  {http://adsabs.harvard.edu/abs/2008MNRAS.388..247D} {388, 247}

\makeatother
\end{thebibliography}

\appendix
\section{Calculation of the foregrounds contribution}
\label{appendix:a}
Here we provide an explicit calculation of the contribution of the foregrounds to the angular power spectrum of Kepler SNR, which are likely to be dominant at smaller angular scales.

As mentioned earlier, Kepler SNR is a weaker source than Cas A, Crab SNR. Hence, the measured $C^E_{\ell}$ at smaller angular scales can be strongly affected by foregrounds like extragalactic point sources and diffuse Galactic synchrotron emission (DGSE). Both of the foregrounds have an inverse frequency scaling. Therefore it is possible that lower frequency L band is more contaminated than C band. We carefully estimate the amount of contribution of the individual foregrounds for the two bands and analyze its effect on the measured $C^E_\ell$. 

We use a mean frequency spectral index of $\alpha =2.50$ \citep{deCosta2008} for diffuse synchrotron emission and calculate the expected amplitude of $C_{\ell}$
\begin{equation}
C_{\ell, \nu}=513mK^{2}\bigg(\frac{1000}{\ell}\bigg)^{2.34}\bigg(\frac{\nu}{150MHz}\bigg)^{-2\alpha}
\end{equation}
\citep{Ghosh2012} using $\nu =1.5$GHz and $\nu =5$GHz for L and C band respectively. A particular value of $\ell$ ($\ell=2.51\times 10^{5}$) is chosen from range III. The amplitude of expected DGSE is $6-7$ order of magnitude lesser than that of the measured $C^E_{\ell}$ at the operating frequency of the bands and for all values of $\ell$. However, the contribution of point sources is independent of $\ell$. We refer to an earlier work \citep{Singal2010} to evaluate the amplitude of $C_{\ell}$ due to discrete point sources. For the flux of the brightest source $S_{c}$ in the FoV, the Poisson contribution of the point sources is given by
\begin{equation}
\label{eqn:C_L_PS}
C_{\ell}= \bigg(\frac{\partial B}{\partial T}\bigg)^{-2} \int\limits_{0}^{S_{c}} S^{2} \frac{dN}{dS}dS.
\end{equation}
The value of $S_c$ considered here is $5\sigma$ where the value of $\sigma$ is taken from a region of the image, outside the source. This $5\sigma$ value is comparable to the maximum flux of the image.
We estimate the $C_{\ell}$ in equation (\ref{eqn:C_L_PS}) using the different source count 
\begin{equation}
\frac{dN}{dS}= k S^{-\gamma}
\end{equation}
from \cite{Ciliegi2003} for C band and \cite{Bondi2003} for L band. The amplitude of $C_\ell$ calculated using eq. (\ref{eqn:C_L_PS}) is less than that of the measured $C^E_{\ell}$ at the second largest $\ell$ value for C band. Considering L band, the point source contribution of $C_{\ell}$  is in excess of $C^E_{\ell}$  for range III. Therefore, the range III of L band may be contaminated by the foregrounds and is excluded from being the characteristic of the remnant.

\bsp	
\label{lastpage}
\end{document}